\documentclass[11pt]{article}
\usepackage{amsmath, amssymb}
\usepackage{verbatim}
\usepackage{latexsym}
\def\be{\begin{eqnarray}}
 \def\ee{\end{eqnarray}}
 \def\0{\nonumber}
\def\d{\partial}
\usepackage{euscript}
\newcommand\EE{\EuScript{E}}
\newcommand\EF{\EuScript{F}}
\usepackage{amsfonts}
\usepackage{verbatim}

\newcommand\E{{\cal E}}

\def\g{{\rm g}}
\def\vareta{\tilde\eta}
\def\ve{\epsilon}
\def\e{\varepsilon}

\def\del{\partial}

\def\del{\partial}

\def\cos{{\rm cos}}
\def\sin{{\rm sin}}

\def\ket#1{|#1 \rangle}
\def\0{\nonumber}

\begin{document}

\begin{flushright}
SISSA/53/2011/EP\\  
hep-th/1109.4336\\
expanded version
\end{flushright}
\vskip 2cm
\begin{center}

{\LARGE {\bf Lump solutions in SFT. Complements}}
 
\vskip 1cm

{\large L. Bonora$^a$, S.Giaccari$^a$ and D.D.Tolla$^b$}

{}~\\

\quad \\

{\em ~$~^{a}$International School for Advanced Studies (SISSA),}\\

{\em  Via Bonomea 265, 34136 Trieste, Italy and INFN, Sezione di Trieste}\\

 {\tt bonora@sissa.it, giaccari@sissa.it}

{}~\\

\quad \\

{\em ~$~^{c}$Department of Physics and University College, }\\

{\em Sungkyunkwan University, }\\

{\em  Suwon 440-746, South Korea}\\

{\tt  dribatolla@gmail.com}
\end{center} 

\vskip 2cm {\bf Abstract.}
{In this paper we complete the analysis started in ArXiv:1105.5926 [hep-th] and ArXiv:1106.3914 [hep-th] where lump solutions were proposed and their energy calculated, and confirm the results found therein. We also rectify the interpretation of the $\ve$ parameter, by showing that it is simply a regulator, not a gauge parameter. Then we closely analyse the would-be violation of the equation of motion for such solutions. We argue that, when the issue is considered in the appropriate mathematical setting, no violations of the equation of motion occur.}

 \eject

\section{Introduction}

This paper is an addition to  \cite{BGT1,BGT2} and a comment to \cite{EM}. Recently, following an earlier suggestion of \cite{Ellwood}, a general method has been proposed, \cite{BMT}, to
obtain new exact analytic solutions in Witten's cubic open string field theory
(OSFT)~\cite{Witten:1985cc}, and in particular solutions that
describe inhomogeneous tachyon condensation. On general grounds it is expected that an OSFT defined on a particular boundary conformal field theory (BCFT) has classical solutions describing other
boundary conformal field theories~\cite{Sen:1999mh,Sen:1999xm}. Analytic solutions have actually been constructed describing the tachyon
vacuum~\cite{Schnabl05, EllwoodSchnabl,Okawa1, ErlerSchnabl, RZ06, ORZ, Fuchs0,
Erler:2006hw, Erler:2006ww, Erler:2007xt, Arroyo:2010fq,
Zeze:2010jv, Zeze:2010sr, Arroyo:2010sy,Murata,Ghoshal} and others describing
general marginal boundary deformations of the initial
BCFT~\cite{KORZ, Schnabl:2007az, Kiermaier:2007vu, Fuchs3,
Lee:2007ns, Kwon:2008ap, Okawa2, Okawa3, Kiermaier:2007ki,
Erler:2007rh}, see also the reviews \cite{Fuchs4, Schnabl:2010tb}. In this panorama
solutions describing inhomogeneous and relevant boundary deformations of the initial BCFT
were not known until recently, though their existence was
predicted~\cite{Sen:1999mh,Sen:1999xm,lumps}. This absence was filled up in \cite{Ellwood,BMT}, and in \cite{BGT1,EM} the energy of a D24-brane solution was calculated for the first time. In \cite{BGT2} 
these results were extended to analytic SFT solutions corresponding to D(25-$p$)-branes, for any $p$, and their energy was calculated. 

Notwithstanding these successes, some formal problems have remained behind. They have given rise to a controversy, as a consequence of which we decided to redo all the (analytic and numerical) calculations of \cite{BGT1,EM}, when possible with different methods and with enhanced computer power. This has taken several months of work, the outcome of which is exposed in this expanded version of a previous preprint. The main results are the following ones: we have found full confirmation of the results in \cite{BGT1,BGT2} and we think we have clarified the controversial issue raised in \cite{EM} (in particular, the Schwinger representation of inverse elements turns out to be flawless if used in the right way). In the process we have corrected a (fortunately innocuous for our final results) misconception present both in \cite{BGT1,BGT2} and in \cite{EM}.

Let us start from the latter point. It concerns the
interpretation of the $\ve$ parameter used in \cite{BGT1,EM,BGT2}. The latter was first introduced in the analysis as a regulator and subsequently  (erroneously) interpreted as a gauge parameter, in the sense that physical quantities were supposed to be independent of it. We show that, both from a theoretical and a numerical point of view, $\ve$ is a mere regulator (not a gauge parameter) and the only meaningful results are obtained in the $\ve\to 0$ limit. This conclusion does not affect the results in \cite{BGT1,BGT2}, because they were obtained precisely in that limit, but it clarifies a theoretical issue which is important in itself and also in relation to the subsequent point.
 
The next problem was raised in \cite{EM} and a solution to it was proposed in \cite{BGT1} in appendix D. In this note we would like to return to this issue and discuss it in full detail. It concerns a would-be violation of the SFT equation of motion for the string field candidates considered in \cite{BMT,BGT1,BGT2,EM}, which originates from the use of a Schwinger parametrization of inverse elements. Our discussion of the problem starts with pointing out that similar problems arise in the search for solutions in classical field theory. We show,
for instance, that were we to take into account terms like the violating term of \cite{EM} (which we call spurious terms) in solving Einstein gravity in vacuum, we would come to the paradoxical conclusion that the Schwarzschild black hole is not a solution of Einstein gravity. We argue that, when the issue is considered in the proper setting, no violations to the equation of motion occur for the solutions considered in \cite{BGT1,BGT2}. The spurious terms when inserted in convergent integrals give vanishing contributions and, on the other hand, can give nonvanishing (but ambiguous) contributions only if inserted in discontinuous integrals
(see below for the precise meaning). This suggests that the appropriate mathematical tool to interpret them is the theory of distributions. We suggest that the lump solutions must be considered as distributions (in the appropriate mathematical setting). Once this is done, any ambiguity linked to spurious terms in the equation of motion, disappears.

The paper is organized as follows. After a review of \cite{BMT} and \cite{BGT1}, section 2 is devoted to the clarification of the nature of the $\ve$ parameter and relevant numerical calculations. In section 3 we outline
the problem that arises when we represent $\frac 1{K+\phi_u}$ by means of a Schwinger parametrization.
In section 4 we discuss in detail the example of the Schwarzschild solution in Einstein gravity and show what would happen if we took into account spurious terms. In section 5
we argue on a general ground that the offending term of \cite{EM} does not have right of
citizenship among well behaved mathematical objects. 
In section 6 we introduce a new formal representation of $\frac 1{K+\phi_u}$, which helps formulating the problem in a clearer way. In section 6 we show that this new representation reproduces the results in
\cite{BGT1}. In section 7 we compute the offending (or spurious) term in the SFT equation of motion and show that it is in fact related to an ambiguity in the formalism, and should not be considered as a matter of principle, but, anyhow, even if taken into account, when inserted in a convergent expression for the energy, this term gives a vanishing contribution. In section 8 we introduces a numerous set of states that can play the role of test states in the distribution theory interpretation of the lump solution.
Section 9 is devoted to a summary of the discussion and results. We also suggest that an appropriate mathematical framework  for the problem discussed in this paper and for similar problems may be based on a re-elaboration of vector distribution theory.

\subsection{Review of the previous results}

In \cite{BMT}, to start with, the well-known $K,B,c$ algebra defined by
\be
K=\frac\pi2K_1^L\ket I, \quad \quad
B=\frac\pi2B_1^L\ket I,\quad\quad c= c\left(\frac12\right)\ket I,
\label{KBc}
\ee
was enlarged as follows. In the sliver frame (obtained by mapping the UHP to
an infinite cylinder $C_2$ of circumference 2, by the sliver map
$\tilde z=\frac2\pi\arctan z$), by adding a (relevant) matter operator
\be \phi=\phi\left(\frac12\right)\ket I\label{phi} \ee with the
properties \be \,[c,\phi]=0,\quad\quad \,[B,\phi]=0,\quad\quad
\,[K,\phi]= \del\phi,\label{proper}
\ee
In this new algebra $Q$ has the following action:
\be
Q\phi=c\del\phi+\del c\delta\phi.
\label{actionQ}
\ee
It can be easily proven that
\be
\psi_{\phi}=c\phi-\frac1{K+\phi}(\phi-\delta\phi) Bc\del c\label{psiphi}
\ee
does indeed satisfy the OSFT equation of motion
\be Q\psi_{\phi}+\psi_{\phi}\psi_{\phi}=0\label{eom}.
\ee
It is clear
that (\ref{psiphi}) is a deformation of the Erler--Schnabl solution,
see \cite{ErlerSchnabl}, which can be recovered for $\phi=1$.

In order to prove that (\ref{psiphi}) is a solution, one demands that
$(c\phi)^2=0$, which requires the OPE of  $\phi$ at nearby points to
be not too singular.

Using the $K,B,c,\phi$ algebra one can show that
\be
{\cal Q}_{\psi_\phi} \frac
B{K+\phi}=Q\frac B{K+\phi}+\left\{\psi_\phi,\frac
B{K+\phi}\right\}=1.\0
\ee
So, unless the homotopy--field $\frac
B{K+\phi}$ is singular, the solution has trivial cohomology, which
is the defining property of the tachyon vacuum
\cite{Ellwood,EllwoodSchnabl}. On the other hand, in order for the
solution to be well defined, the quantity
$\frac1{K+\phi}(\phi-\delta\phi)$ should be well defined too. Finally,
in order to be able to show that (\ref{psiphi}) satisfies the
equation of motion, one needs $K+\phi$ to be invertible.

In full generality we thus have a new nontrivial solution if
\begin{enumerate}
\item $\frac 1{K+\phi}$ is singular, but
\item $\frac 1{K+\phi}(\phi-\delta\phi)$ is regular and
\item $\frac 1{K+\phi}(K+\phi)=1.$
\end{enumerate}

In \cite{BMT} sufficient conditions for $\phi$ to comply with
the first two requirements were determined. Let us parametrize the
worldsheet RG flow, referred to above, by a parameter $u$, where
$u=0$ represents the UV and $u=\infty$ the IR, and rewrite $\phi$ as
$\phi_u$, with $\phi_{u=0}=0$. Then we require for $\phi_u$ the
following properties under the coordinate rescaling $f_t(z)=\frac zt$
\be
f_t\circ\phi_u(z)=\frac 1t\,\phi_{tu}\left(\frac
zt\right)\label{cnd1}
\ee
and, most important, that the partition function
\be
g(u)\equiv Tr[e^{-(K+\phi_u)}]=\left\langle
e^{-\int_0^1ds\, \phi_{u}(s) }\right\rangle_{C_1},\label{g(u)Tr}
\ee
satisfies the asymptotic finiteness condition
\be
\lim_{u\to\infty}\left\langle e^{-\int_0^1ds\, \phi_u(s)
}\right\rangle_{C_1}=\verb"finite".\label{cnd3}
\ee
It was pointed out in \cite{BMT} that this satisfies the first two conditions above i.e.
guarantees not only the regularity of the solution but also its
'non-triviality', in the sense that if this condition is satisfied,
it cannot fall in the same class as the ES tachyon vacuum solution.
It would seem that the last condition above cannot be satisfied in
view of the first. But this is not the case. This is the main issue discussed in sec.3,5-8.

We will consider in the sequel a specific relevant operator $\phi_u$
and the corresponding SFT solution. This operator generates an exact
RG flow studied by Witten in \cite{Witten}{},  see also
\cite{Kutasov}{}, and is based on the operator (defined in the
cylinder $C_T$ of width $T$ in the arctan frame)
\be
\phi_u(s) = u(X^2(s)+2\ln u +2 A)\label{TuCT1},
\ee
where $A$ is a constant first
introduced in \cite{Ellwood}. In $C_1$  we have
\be
\phi_u(s) = u
(X^2(s)+ 2 \ln Tu +2 A)\label{TuC1b1} \ee and on the unit disk $D$,
\be \phi_u(\theta) = u (X^2(\theta)+ 2\ln \frac {Tu}{2\pi} +2
A).\label{TuDb1}
\ee

If we set
\be
g_{A}(u)= \langle e^{-\int_0^1ds \,
\phi_u(s)}\rangle_{C_1} \label{gAu1} \ee we have \be g_{A}(u)
=\langle e^{- \frac 1{2\pi } \int_{0}^{2{\pi}} d\theta \, u\Bigl{(}
X^2(\theta) + 2 \ln \frac u{2\pi}+2 A\Bigr{)}}\rangle_{D}.  \0
\ee
According to \cite{Witten},
\be
g_{A}(u) = Z(2u)e^{-2u (\ln \frac
u{2\pi}+A)},\label{gAu2} \ee where \be Z(u)= \frac 1{\sqrt{2\pi}}
\sqrt{u} \Gamma(u)e^{\gamma u}\label{Z(u)}
\ee

Requiring finiteness for $u\to\infty$ we get $A= \gamma -1+\ln
4\pi$, which implies
\be
g_{A}(u)\equiv g(u)= \frac 1{\sqrt{2\pi}}
\sqrt{2u} \Gamma(2u) e^{2u(1-\ln (2u))}\label{gA1unorm}
\ee
and
\be
\lim_{u\to\infty} g(u) = 1.\label{inflimitgA1u} \ee
Moreover, as it turns out, $\delta \phi_u=- 2u$, and so: \be \phi_u -\delta\phi_u =
u \d_u \phi_u(s).\label{uduphiu1}
\ee
Therefore the $\phi_u$ just introduced satisfies all the required properties and consequently
$\psi_u\equiv \psi_{\phi_u}$ must represent a D24 brane solution.

In I the expression for the energy of the lump solution was determined by evaluating a three--point
function on the cylinder $C_T$ of circumference $T$ in the arctan frame. It is given by
\be
E[\psi_u]&=&-\frac16 \langle \psi_u \psi_u\psi_u\rangle\0\\
&=&\frac16 \int_0^\infty d(2uT)\; (2uT)^2\int_0^1
dy\int_0^{y} dx\,\frac4\pi \,\sin\pi x\,\sin\pi y\,\sin\pi(x-y)\label{Efinal}\\
&&\cdot g(uT)\Bigg\{-\Big(\frac{\partial_{2uT}g(uT)}{g(uT)}\Big)^3
+G_{2uT}(2\pi x)G_{2uT}(2\pi(x-y))G_{2uT}(2\pi y)\0\\
&&-\frac 12 \Big(\frac{\partial_{2uT}g(uT)}{g(uT)}\Big)\Big(G_{2uT}^2(2\pi
x)+G_{2uT}^2(2\pi(x-y))+G_{2uT}^2(2\pi y)\Big) \Bigg\}.\0
\ee
where $G_u(\theta)$ represents the
correlator on the boundary, first determined by Witten, \cite{Witten}:
\be
G_u(\theta)= \frac {1}{u} +2 \sum_{k=1}^\infty
\frac {\cos (k\theta)}{k+u} \label{Gutheta}
\ee
Moreover ${\cal E}_0(t_1,t_2,t_3)$ represents the ghost three--point function in $C_T$.
\be
\E_0(t_1,t_2,t_3)=\left\langle Bc\d c(t_1+t_2)\d c(t_1) \d c(0)
\right\rangle_{C_T} = -\frac 4{\pi} \sin \frac {\pi t_1}T \sin \frac
{\pi(t_1+t_2)}T \sin \frac{\pi t_2}T.\label{E0}
\ee
Finally, to get (\ref{Efinal}) a change of variables $(t_1,t_2,t_3)\to (T,x,y)$, where \be x=\frac{t_2}{T},
\quad\quad y=1-\frac{t_1}{T}.\0
\ee
is needed.

The expression (\ref{Efinal}) has been evaluated in \cite{BGT1}. As it turns out, this expression has
a UV ($s\approx 0$, setting $s=2uT$) singularity, which must be subtracted away. Therefore the result one
obtains in general will depend on this subtraction\footnote{The subtraction does not fix by itself the zero-point energy. For instance, in the examples of \cite{BGT2}, the expression corresponding to (\ref{Efinal}) is explicitly gauge dependent.}. In \cite{BGT1} it has been pointed out that {\it a physical
significance can be assigned only to a subtraction-independent quantity}, and it has been shown how to define and evaluate such a quantity. First a new solution to the EOM, depending on a regulator $\ve$, has been introduced\footnote{In \cite{BGT1} $\psi_u^{\ve}$ was called $\psi_\ve$.}
\be
\psi_u^{\ve}= c(\phi_u+\ve) - \frac 1{K+\phi_u+\ve} (\phi_u+\ve -\delta \phi_u) Bc\partial c.
\label{psiphieps}
\ee
its energy being 0 (after the same UV subtraction as in the previous case) in the $\ve\to 0$ limit.
Then, using it, a solution to the EOM at the tachyon condensation vacuum has been obtained. The equation of motion at the tachyon vacuum is
\begin{align}\label{EOMTV}
{\cal Q}\Phi+\Phi\Phi=0,\quad {\rm where}~~{\cal Q}\Phi=Q\Phi+\psi_u^\ve\Phi+\Phi\psi_u^\ve.
\end{align}
One can easily show that
\begin{align}\label{psiupsie}
\Phi_0^\ve=\psi_u-\psi_u^\ve
\end{align}
is a solution to (\ref{EOMTV}). The action at the tachyon vacuum is $-\frac12\langle{\cal
Q}\Phi,\Phi\rangle-\frac13\langle\Phi,\Phi\Phi\rangle.$ Thus the energy of of the lump, $E[\Phi_0]$, is
\be\label{TVlumpenergy}
E[\Phi_0]&=&-\lim_{\ve\to 0}\frac16\langle\Phi_0^\ve,\Phi_0^\ve\Phi_0^\ve\rangle\0\\
&=&
-\frac16\lim_{\e\to 0} \big[\langle\psi_u,\psi_u\psi_u\rangle-\langle\psi_u^\ve,\psi_u^\ve\psi_u^\ve\rangle
-3\langle\psi_u^\ve,\psi_u\psi_u\rangle+3\langle\psi_u,\psi_u^\ve\psi_u^\ve\rangle\big].
\ee
The integrals in the four correlators at the RHS, are IR ($  s\to\infty$) convergent.
The UV subtractions necessary for each correlator are always the same, therefore they cancel out. In \cite{BGT1}, after UV subtraction, we obtained
\be
 &&-\frac 16 \langle\psi_u,\psi_u\psi_u\rangle=\alpha+\beta, \quad\quad \lim_{\ve\to 0}\langle\psi_u^\ve,\psi_u^\ve\psi_u^\ve\rangle=0\0\\
&&\frac 16 \lim_{\ve\to 0}\langle\psi_u^\ve,\psi_u\psi_u\rangle=\alpha-\frac 23 \beta,
\quad\quad   \frac 16 \lim_{\ve\to 0} \langle\psi_u,\psi_u^\ve\psi_u^\ve\rangle=
\alpha -\frac 13 \beta\label{finalBGT1}
\ee
where $\alpha+\beta\approx 0.068925$\footnote{This number represents the result of an improved numerical evaluation and differs from the value given in \cite{BGT1} by 6 per mil.} was evaluated numerically and $\alpha= \frac 1{2\pi^2}$ was calculated analytically. So $E[\Phi_0]= \alpha$ turns out to be precisely the D24-brane energy. In \cite{BGT2} the same result was extended to any Dp-brane lump.

\section{Nature of the $\ve$ parameter}

Eq.(\ref{TVlumpenergy}) and (\ref{finalBGT1}) is really what we proved in \cite{BGT1,BGT2}. That is, the result we obtained is only valid in the limit $\ve \to 0$, and this was the correct thing to do. However we were mislead by a wrong theoretical prejudice and by a too rough numerical result
into believing that the expression in square brackets in the RHS of (\ref{TVlumpenergy}) is independent
of $\ve$, and therefore $\ve$ can be interpreted as a gauge parameter. This is not the case, as we will show in this section: {\it $\ve$ is a simple regulator} and physical quantities can be recovered only in the $\ve\to 0$ limit.

In \cite{BGT1} we computed numerically $\langle\psi_u^\ve,\psi_u^\ve\psi_u^\ve\rangle$, after making the necessary UV subtraction. The result was reported in Table 3 there, and led us to the idea that that is evidence of the analytic result being 0 for any $\ve$. This convinced us that $\ve$ is a gauge parameter and, as a consequence, also the full expression in square brackets in the RHS of (\ref{TVlumpenergy}) should not
depend on $\ve$. Although it did not have any practical consequence on the final result, it must be said that  this is not true. The present section is devoted to clarifying this issue.

Let us deal first with $\langle\psi_u,\psi_u\psi_u\rangle-\langle\psi_u^\ve,\psi_u^\ve\psi_u^\ve\rangle$.
One of the limits of the numerical evaluation of $\langle\psi_u^\ve,\psi_u^\ve\psi_u^\ve\rangle$ in \cite{BGT1} 
was that the numerics can start only after the UV subtraction is carried out. This limits considerably the accuracy of the numerical approximation. The expression
\be
\Delta^{(1)}_\ve= \langle\psi_u,\psi_u\psi_u\rangle-\langle\psi_u^\ve,\psi_u^\ve\psi_u^\ve\rangle,\0
\ee 
instead, is UV finite  
and its numerical evaluation can be more accurate. The analytical preparation for this evaluation is in
Appendix A. Here we report the numerical results for a sample of values of the parameter $\eta=\frac {\ve}{2 u}$.
\begin{table}[ht]
$\begin{matrix}\quad &\eta:  &2\,&1 \, &  0.7\,& 0.5\,& 0.1\,&0.08\\
\quad&\Delta^{(1)}_\ve :\quad&  -0.41968\quad& -0.41958\quad&-0.42028\quad&  -0.41860\quad& -0.41868\, \quad& -0.41853\\
{}&{}&{}&{}&{}&{}&{}&{}\\
\quad &\eta: &0.05 \, &  0.01\,& 0.005\,&\,0.003\,&\,0.001&\,0.0005\\
\quad&\Delta^{(1)}_\ve :  \quad& -0.41831\quad&-0.41660\quad& -0.41625\quad&-0.41587\quad&- 0.41483\quad&- 0.414009\quad\\
\end{matrix}$
\caption{Samples of $ \Delta^{(1)}_\ve$ }
\end{table}

The limit $\lim_{\ve\to 0}\Delta^{(1)}_\ve $ was calculated in \cite{BGT1} and is given by: $ 6(\alpha+\beta)\approx - 0.41355$. Since the numbers in Table 1 are accurate up to the third digit (being very conservative the error can be estimated to be $\pm 0.0005$) the dependence on $\ve$ is evident. It is also clearly visible that the sequence of numbers tends to the expected value (around $\eta = 0.00001$ reliable numerical results becomes hard to retrieve). The smallness of the $\ve$ dependence  (a few percent only) was at the origin of the misunderstanding about the nature of $\ve$.

The dependence on $\ve$ of 
\be
\Delta^{(2)}_\ve= \langle\psi_u^\ve,\psi_u\psi_u\rangle-\langle\psi_u,\psi^{\ve}_u\psi^{\ve}_u\rangle\0
\ee 
is not much easier to detect. In Appendix A one can find the preliminaries to the numerical calculations. In Table 2 we report the numerical results for a sample of the parameter $\eta$.

\begin{table}[ht]
$\begin{matrix}\quad &\eta:  & 10\,&2\,&1 \, &  0.7\,& 0.5\,\\
\quad&\Delta^{(2)}_\ve : \quad& -0.01431 \quad&   -0.02704\quad& -0.0308524\quad&  -0.0323693\quad& -0.03332\\
{}&{}&{}&{}&{}&{}&{}&{}\\
\quad &\eta:&0.4\,  \,&0.2\,&0.1 \, &  0.08\,& 0.05\,\\
\quad&\Delta^{(2)}_\ve :\quad&-0.03398 \quad&  - 0.03525\quad& -0.03567\quad&-0.03550\quad&  -0.03613\\
\end{matrix}$
\caption{Samples of $ \Delta^{(2)}_\ve$ }
\end{table}

In \cite{BGT1} the numerical value of $\Delta^{(2)}_\ve$ was determined in the $\ve\to 0$ limit to be: $\lim_{\ve\to 0} \Delta^{(2)}_\ve= -2\beta \approx -0.03652$.
The results in Table 1 are to be taken with a possible uncertainty of $\pm 0.0005$. We see that they clearly depend on $\ve$ and that the limit $\ve\to 0$ tends to the expected value.

After these results the dependence on $\ve$ of (\ref{TVlumpenergy}) needs not be stressed. {\it $\ve$ is no gauge parameter}, it is a simple regulator, as it was originally conceived. This conclusion could have been
reached from a theoretical point of view. We see in fact that, while our $\phi_u(s)$ satisfies condition 
(\ref{cnd1}), the combination  $\phi_u(s)+\ve$, which appears in $\psi_u^\ve$, does not since:
\be
f_t\circ \ve = \ve \neq \frac {\ve}t \0
\ee
Forcing $\ve$ to satisfy (\ref{cnd1}) would require $\ve=\kappa u$ for some positive constant $\kappa$; but then, in $\phi_u+\ve$, see (\ref{TuCT1}), $\ve$ could be absorbed into a redefinition of $A$ and would disappear from $\phi_u^\ve$. As a consequence the latter would actually coincide with $\psi_u$ and $\Delta_\ve^{(1)}$ would vanish, which is evidently not the case. The role of $\ve$ is precisely to break the covariance under
the semigroup of rescalings, eq.(\ref{cnd1}), in order to generate a different kind of solution with respect to $\psi_u$. The conclusion is that the parameter $\ve$ does not run (in the RG parlance), therefore it is not a gauge parameter (in the SFT terminology). We remark that the value $\ve=0$ is (together with $\ve=\infty$) the
only scale invariant one.

One may be surprised at first that $\psi_u^\ve$ is a solution to the EOM of SFT, while the
term $ \langle\psi_u^\ve,\psi_u^\ve\psi_u^\ve\rangle$ is $\ve$-dependent . The point is that $\psi_u^\ve$ 
formally solves the equation of motion but is not an extreme of the action for $\ve\neq 0$. The puzzle is explained of course by the fact that the parameter $\ve$ is not present in the original action. Therefore one has to prove 
{\it a posteriori} that the `solution' actually corresponds to an extreme of the action\footnote{The same consideration applies also to the parameter $u$, but it was shown in \cite{BMT} that $u$ actually disappears from the action when we replace $\psi_u$ in it: $u$ is a true gauge parameter.}.
The variation of the action with $\ve$ is given by (after replacing the eom) $\delta_{\ve}S\sim \langle \frac {\del \psi_u^\ve} {\del {\ve}}, Q \psi_u^\ve\rangle -
\langle Q \frac {\del \psi_u^\ve} {\del {\ve}}, \psi_u^\ve\rangle$. For this to vanish one should be able to `integrate by parts', which is not possible due to the UV subtractions implicit in the calculation of the correlators, see \cite{BGT1} (and also \cite{Hata} where similar arguments are developed although not in the same context)\footnote{Since the UV singularity is linked to the $X$ zero mode, one might expect that with a compactified $X$ this problem should disappear and the integration by parts become possible. However, as long as we consider solution of the type $\psi_u,\psi_u^\ve$ with a linearly scaling $u$ parameter, this seems to be impossible: the singularity removed from the UV will pop up in the IR, creating analogous problems. The nontrivial boundary contribution in the SFT action, see also section 2, is a new interesting feature which deserves a closer investigation.}. Now $\delta_{\ve}S$ does not vanish and in order to find an extreme of the action we have to extremize it. This is in keeping with the monotonic dependence on $\eta$ one can see in Table 1, which tells us that the extreme is met in the limit $\ve\to 0$.

We have verified that also other quantities considered in section 7 of this paper, which contain $\ve$, are effectively $\ve$-dependent.
In the light of the above theoretical argument, this and the previous numerical proof that $\Delta_\ve^{(1)}$ and $\Delta_\ve^{(2)}$ are $\ve$-dependent would be pointless, if a misunderstanding
about the role of $\ve$ had not arisen. In any case, having at hand Table 1 and 2, we have the opportunity to make the following observation: the limit $\ve\to 0$ is smooth and it tends 
to the expected theoretical value (just replace the numerical values of $\Delta_\ve^{(1)}$ and $\Delta_\ve^{(2)}$ inside (\ref{TVlumpenergy})). Nothing anomalous happens in the limit. We have proven the existence of the limit also analytically (we will spare the reader the lengthy details), but the numerical results are more pictorial. Eq.(\ref{TVlumpenergy}) was obtained by plugging in ${\cal Q}\Phi_0=- \Phi_0\Phi_0$ into the SFT action. Should the EOM be violated, the worst that can happen is that the violating term, if any, contributes 0 to the energy. This is exactly what we will show in section 7.

\section{The problem with the Schwinger representation}

We now come to the criticism raised by \cite{EM} about our solution. 
In order to obtain (\ref{Efinal}) one has to use the following Schwinger representation
\be
\frac 1{K+\phi_u}= \int_0^\infty dt\, e^{-t(K+\phi_u)}\label{Schright}
\ee
of the inverse of $K+\phi_u$. When using such a Schwinger representation, however, the identity
\begin{align}\label{IDDu}
\frac1{K+\phi_u}(K+\phi_u)=I,
\end{align}
would seem not to be satisfied. To illustrate the problem, let us
calculate the overlap of both the left and the right hand sides of
\eqref{IDDu} with $Y=\frac12\partial^2c\partial cc$. The right hand
side is trivial and, in our normalization, it is
\begin{align}\label{LHS}
{\rm Tr}(Y\cdot I)=\lim_{t\to0}\langle Y(t)\rangle_{C_t}\langle
1\rangle_{C_t}=\frac V{2\pi}.
\end{align}
To calculate the left hand side we need the Schwinger representation
\begin{align}\label{RHS}
{\rm Tr}\big[Y\cdot
\frac1{K+\phi_u}(K+\phi_u)\big]=\int_0^{\infty}dt{\rm Tr}\big[Y\cdot
e^{-t(K+\phi_u)}(K+\phi_u)\big]
\end{align}
Making the replacement
\begin{align}\label{naive}
e^{-t(K+\phi_u)}(K+\phi_u)\to -\frac d{dt}e^{-t(K+\phi_u)}
\end{align}
one obtains
\begin{align}\label{wrong}
{\rm Tr}\big[Y\cdot
\frac1{K+\phi_u}(K+\phi_u)\big]=g(0)-g(\infty)=\frac
V{2\pi}-g(\infty),
\end{align}
which is different form \eqref{LHS} because $g(\infty)$ is nonvanishing.
The latter relation is often written in a stronger form
\be
\int_0^\infty dt\, e^{-t(K+\phi_u)}(K+\phi_u)= 1- \Omega_u^\infty, \quad\quad  \Omega_u^\infty=\lim_{\Lambda\to\infty} e^{-\Lambda(K+\phi_u)}\label{1-Omega}
\ee
This (strong) equality, however, has to be handled with great care. If the latter is taken literally, we could also write
\be
\frac 1{K+\phi_u}= \int_0^\infty dt\, e^{-t(K+\phi_u)}+ \frac 1{K+\phi_u} \Omega_u^\infty
\label{Schwrong}
\ee
instead of (\ref{Schright}). This would imply that eq.(\ref{IDDu})
is not satisfied, and, consequently, the equation of motion is not satisfied by $\psi_u$.

\section{An example from classical field theory}

The problem raised in the previous section is actually commonplace in the search of solutions in ordinary 
classical field theory and was solved long ago resorting to the theory of distributions, and tacitly incorporated in our common lore. Let us present one of many possible examples. We ask our patient reader to follow us through some elementary mathematics. We hope the example will help clarifying our line of thought in solving the puzzle raised in the previous section.

\subsection{Preliminaries}

In preparation for our 3d example let us introduce some notation:
\be
x&=& r \sin \theta \cos \varphi, \quad y=r \sin\theta \sin \varphi, \quad z=r\cos \theta\0\\
r&=&\sqrt{x^2+y^2+z^2}, \quad \theta = \arccos \frac zr ,\quad \varphi= \arctan \frac yx \0
\ee
and
\be
\Delta f = \frac 1{r^2} \partial_r \left(r^2 \partial_r f\right) +\frac 1{r^2\sin\theta } \partial_\theta
\left(\sin\theta\, \partial_\theta f\right) + \frac 1{r^2 \sin^2\theta} \partial_\varphi^2 f\label{laplacian}
\ee
Distribution theory tells us that
\be 
\Delta \frac 1r = - 4\pi \delta(r)\label{Delta1r}
\ee
Now, let us consider the product $r \frac 1r$. According to distribution theory (and to continuity) we should have
\be 
r \frac 1r=1\label{right}
\ee
In fact, using a test function $f(x,y,z)$, we have
\be
&&<r \frac 1r,f>= \lim_{\ve\to 0}{\int\int\int}_{r\geq \ve}dxdydz\, r \frac 1r\, f(x,y,z)\0\\
&=& \lim_{\ve\to 0}{\int\int\int}_{r\geq \ve}dxdydz\, f(x,y,z) = {\int\int\int} dxdydz\, f(x,y,z)= <1, f>\0
\ee
Therefore $r \frac 1r$, as a distribution, is 1. In view of the previous section one might decide to use a Schwinger representation
\be 
r \frac 1r\longrightarrow \int_0^\infty dt\, r e^{-tr} = -  \int_0^\infty dt \frac {\partial}{\partial t} e^{-tr} =
1- \lim_{t\to \infty} e^{- t r}\equiv 1-\Omega(r)\label{r1rschwinger}
\ee
Consequently $\frac 1r$ is represented by
\be
\frac 1r \longrightarrow \int_0^\infty dt\, e^{-tr} + \Pi(r),\quad\quad \Pi(r)=\frac{ \Omega(r)}r\label{schwinger2}
\ee

Let us elaborate a bit on this in order to prepare the ground for our example. From (\ref{r1rschwinger})
it is clear that $\Omega(r)$ has support at $r=0$, therefore it must be a delta-function-like object. 
Using the definition of delta function as a limit 
\be 
\lim_{t\to \infty} \sqrt{\frac t\pi } e^{-t x^2}= \delta(x),\0
\ee 
we can set
\be 
\Omega(r) =2 \lim_{t\to\infty} \sqrt {\frac {\pi r}t} \delta(r) \label{approxdelta}
\ee
This is the only way to bring $\Omega(r)$ in the world of well-defined objects.
(\ref{approxdelta}) is obviously an identically vanishing distribution. But, of course, if we integrate it over 
something which is not a test function, we may get a nonvanishing result.
For later use let us define also
\be 
\Pi(r) = 2 \lim_{t\to\infty} \sqrt {\frac {\pi }{rt}} \delta(r), \quad\quad
 \Xi(r) = 2 \lim_{t\to\infty} \sqrt {{\pi }{rt}} \delta(r)\label{approxdelta3}
\ee

Let us make a comparison between (\ref{right}) and (\ref{r1rschwinger}). An explicit calculation yields
\be 
\Delta \left( r \frac 1r\right) = - 4\pi r \delta(r) + \frac 2 {r^2}-\frac 2 {r^2} = 0\label{right2}
\ee
as a distribution. This is a result of (\ref{Delta1r}), of $\Delta r= \frac 2r$ and of
\be 
\sum_{i=1}^3 \partial_{x_i} \frac 1r \, \partial_{x_i} r = - \frac 1 {r^2}\label{dxidxi}
\ee
This last calculation is straightforward for $r\neq 0$, but at the origin one must be careful and use distribution theory: for a test function $f(x,y,z)$ we can write, for example,
\be
&&{\int\int\int}dxdydz \,\partial_x\left(\frac 1r \right)  f(x,y,z)=-\lim_{\e\to 0} {\int\int\int}_{r\geq \ve}dxdydz
 \,\sin\theta\, \cos \varphi \frac 1{r^2}\,f(x,y,z) \0\\
&&-\lim_{\e\to 0} {\int\int\int}_{r\geq \ve}dr d\theta d\varphi \, \sin^2\theta\,\cos \varphi f(r,\theta,\varphi)= - {\int\int\int}d r d\theta d\varphi \, \sin^2\theta\,\cos \varphi f(r,\theta,\varphi)\0
\ee
which means that the distributional derivative $\partial_x \frac 1r$ coincides with the ordinary derivative
(there is no extra contribution from $r=0$).

On the other hand, using the representation (\ref{r1rschwinger}), we have 
\be
\Delta \left( r \frac 1r\right) = -\Delta \Omega(r)\label{wrong1}
\ee
The RHS is formally nonvanishing since $\Delta \left(\sqrt r \delta(r)\right) = \frac 34 r^{-\frac 32}\delta(r)
+ 3 r^{-\frac 12}\delta'(r)+ r^{\frac 12}\delta''(r)$. However, remembering that the volume element contains
a factor of $r^2$, $\Delta \Omega(r)$ is in fact the 0 distribution. This is consistent with (\ref{right2}).
But if we do not correctly apply the rules of distribution theory the RHS of (\ref{wrong1}) may seem to be nonvanishing (although ambiguous). This may happen, for instance, if we integrate such term multiplied by a function
that is more singular than $\frac 1{\sqrt r}$ for $r\approx 0$. The trouble is that such a function is not a test function.

\subsection{The Schwarzschild black hole `non-solution'}

Let us check on an example that the (wrong) use of (\ref{wrong1}) leads to wrong results. To this end
we consider the Schwarzschild solution in gravity. The Schwarzschild geometry is a solution to the Einstein
equation in vacuum: $R_{\mu\nu}=0$.

Let us consider the ordinary approach.  The Schwarzschild metric has the form
\be 
ds^2= -f(r)dt^2 +\frac {dr^2}{f(r)} + r^2 (d\theta^2+\sin^2 \theta d\varphi^2)\label{Schw1}
\ee
so that we have 
\be 
g_{00} = -f(r),\quad\quad g_{rr} = \frac 1{f(r)},\quad\quad g_{\theta\theta}= r^2,\quad\quad
g_{\varphi\varphi}= r^2 \sin^2\theta\0
\ee
where, for simplicity, we take $f(r)=1-\frac {2M}r$. The Christoffel symbols are
\be 
\Gamma^0_{0r}= \frac {f'}{2f}=-\Gamma^r_{rr}, \quad\quad\Gamma^r_{00}=\frac {ff'}2, \quad\quad\Gamma^r_{\theta\theta}= -rf,\quad\quad \Gamma^r_{\varphi\varphi}= -rf \sin^2\theta,\label{Christ}
\ee
where $f'(r)= \frac {df(r)}{dr}$. There are other (completely angular) nonvanishing symbols but we will not need them. As a consequence in particular we have
\be 
R_{0r0r}=\frac {f''}2,\quad\quad R_{0\theta 0\theta}= \frac 12 rff', \quad\quad  R_{0\varphi 0\varphi}=
\frac 12 rff' \sin^2\theta\label{Riemann}
\ee
At this point it is easy to prove, for instance, that
\be 
R_{00}= g^{rr} R_{0r0r}+g^{\theta\theta}R_{0\theta 0\theta}+ g^{\varphi\varphi} R_{0\varphi 0\varphi}=0\label{Ricci}
\ee
so that the eom is satisfied (for the $_{00}$ case). 

In all the above, $\frac 1f$ is singular at the horizon $r=2M$, so that one component of the metric is singular. However the Riemann tensor is not singular (and the energy is finite).  In the intermediate passages we have to manipulate $\frac 1f$ or derivative thereof. This is singular, but interpreting it and carrying out all the operations in the framework of distribution theory all the singularities can be treated correctly and the final result is regular.

Now let us see what happens instead when we use improperly the Schwinger representation for $\frac 1f$. To this end let us call
\be 
S(\frac 1f)= \int_0^\infty dt\, e^{-t f}\label{S1f}
\ee
the Schwinger representation of $\frac 1f$. We have
\be
&&f S(\frac 1f) = 1- \Omega (f),\quad\quad
\frac 1f= S(\frac 1f)+ \Pi(f)\0\\
&& f \Pi(f)= \Omega(f),\quad\quad
f\, \partial_r S(\frac 1f)= f' \Xi(f) -f' S(\frac 1f)\label{relations}
\ee
The last one follows from
\be 
f \,\partial_r S(\frac 1f)&=& - \int_0^\infty dt\, t\, f' f e^{-tf} = f'\int_0^\infty dt\,t \frac {d}{dt} e^{-tf}\label{fdrSf}\\
&=& f' \int_0^\infty dt \,\frac {d}{dt}\left(t \,e^{-tf}\right) - f'   \int_0^\infty dt\, e^{-tf}=
2f' \lim_{t\to\infty} \sqrt{\pi t f} \delta(f) - f'  S(\frac 1f)\0
\ee
Of the relevant Christoffel symbols $\Gamma_{0r}^0, \Gamma_{rr}^r$ are singular, while the others are regular.
More precisely we have
\be 
\Gamma_{0r}^0= \frac {f'}2 \left(S(\frac 1f)+\Pi(f)\right),
\quad\quad \Gamma_{rr}^r=  \frac {f'}2 \left( \Xi(f)- S(\frac 1f)\right)\label{Christsing}
\ee
Repeating the calculation with these inputs we get
\be 
R_{0r0r} &=& \frac {f''}2 +  \frac {ff'}2\Pi'(f) +   \frac{f'f'}4 \left(\Xi(f) +\Omega(f) S(\frac 1f)+\Omega(f)\Pi(f)\right)\label{curv}\\ 
R_{0\theta 0\theta}&=& \frac 12 rff',\quad\quad  R_{0\varphi 0\varphi}=
\frac 12 rff' \sin^2\theta\0
\ee
Therefore
\be
R_{00} &=& \frac {ff''}2 + \frac {ff'}{r} \label{R00}\\
&&+ \frac {ff'f'}4 \Xi(f) +\frac {f'f'}4\Omega(f) +  \frac {f^2f'}2 \Pi'(f) \0
\ee
The first line is the usual (vanishing) result, the second line represents the violation to the eom. Notice that {\it in the framework of distribution theory} the second line vanishes, but if one takes the previous algebraic manipulations literally one might conclude that Schwarzschild's is not a solution of Einstein gravity. In 
particular if we integrate the second line over a non-test function we may get something different from 0. This is no accident: these terms are intrinsically ambiguous, as is evident if one tries to define them carefully. Terms such as those in the second line of (\ref{R00}) are inevitably ambiguous when considered outside the framework of distribution theory. We will refer to them as {\it spurious terms}.

This is an example of what we run into when we abandon the principle
of continuity (or analyticity) according to which the statement: $r \frac 1r=1$ everywhere, is the correct thing. This principle has been incorporated into the theory of distributions, which, in this way, has eliminated all the above ambiguities (a distribution is defined via Riemann integrals, which in turn are defined by means of continuous limiting processes, so they automatically incorporate the principle of continuity). But if we abandon this principle we end up in a jungle of contradictions.

\section{Continuity and the Schwinger representation}

The previous example may sound somewhat exotic, but in every respect it is a paradigm of the problem introduced in section 3. Let us now return to it.

In our approach in \cite{BGT1,BGT2} we have always been guided by what we have called above the principle of continuity. On the basis of this principle (\ref{Schright}), as opposed to (\ref{Schwrong}), is the correct relation.
Let us summarize how we discussed this issue  in Appendix D of \cite{BGT1}.
We start from the observation
that $K+\phi_u$ is a vector in an infinite dimensional space: $K+\phi_u=(K_1^L+\phi_u(\frac 12))|I\rangle$, where $|I\rangle$ is the identity string field (and we remark that in our applications $\phi_u(\tilde z)$ is always inserted in the left part of the string). Therefore the inverse of $K+\phi_u$ can also be obtained
via the inverse of the operator $ {\cal K}_u\equiv K_1^L+\phi_u(\frac 12)$.

The operator ${\cal K}_u$ is self-adjoint. Therefore its spectrum lies on the real axis. To know more about it we would need a spectral analysis of ${\cal K}_u$, similar to what has been done for the operator $K_1^L$ in \cite{RSZ,BMST1,BMST2,BMT3}. The spectrum of the latter is the entire real axis. The spectrum of ${\cal K}_u $ is of course expected to be different, but we know on a general ground that it lies on the real axis. We can therefore define the resolvent
of ${\cal K}_u$, $R(\kappa, {\cal K}_u)$, which is by definition the inverse of $\kappa- {\cal K}_u$ ($\kappa$ being a complex parameter). The resolvent is well defined (at least) for any non-real $\kappa$.
We do not know what type of eigenvalue the $\kappa=0$ one is: discrete, continuous or residual. However, since $R(\kappa, {\cal K}_u) (\kappa- {\cal K}_u)=1$ is true for any $\kappa$
outside the real axis, we can hold it valid also in the limit $\kappa\to 0$ by continuity.
Therefore we conclude that, on the basis of the (healthy) principle of continuity,
(\ref{Schright}) is the correct relation, much as $r\frac 1r=1$ was held true everywhere in the previous section. 

The obvious difference between the two cases is that in the previous section's case we were
talking about the inverse of a position $r$, while in this section we are talking about the inverse of a string field  $K+\phi_u$. We remark however that this is the natural correspondence when we pass from classical gravity (classical field theory) to SFT: the role
of positions in the former is played by string configurations in the latter.

One may object at this point that, true, since (\ref{Schright}) is correct, the SFT equation of motion is satisfied by our solution, but in order to compute its energy  we need the Schwinger representation of the inverse of $K+\phi_u$. Given the ambiguity of the latter
(see (\ref{1-Omega}) and (\ref{Schwrong})) brought about by the term $\Omega_u^\infty$, one may wonder whether the computation of the energy may be altered by the presence of such terms.

On the basis of the analogy with the previous section we are led to conclude that such ambiguous terms have to be identified as spurious ones. We have argued above that a good hygienic rule is to drop them. Keeping them may be useless in the best case and misleading in the worst. In any case we would like to modestly
remark that, should we find that the Schwinger representation is defective in calculating the energy, the most logical course would be to correct it, not to blame the solution for not
satisfying the equation of motion. Fortunately, anyhow, this will not be necessary. 
The Schwinger representation perfectly does its job, provided it is handled with care. In fact we will show that spurious terms yield vanishing contributions if inserted in converging integrals, while they may give nonvanishing (but ambiguous) contributions only if they appear in divergent integrands.

\section{A new (formal) representation for $\frac 1{K+\phi_u}$ }

Let us introduce a small nonnegative parameter $\e$, and remark that we can formally write
\be
\frac 1{K+\phi_u}= \frac 1{K+\phi_u+\e-\e}= \sum_{n=0}^\infty \frac {\e^n}{(K+\phi_u+\e)^{n+1}}
\label{infepsisum}
\ee
We can also rewrite it as
\be
\frac 1{K+\phi_u}= \sum_{n=0}^\infty \frac {(-\e)^n}{n!}\,\partial_\e^n \frac 1{(K+\phi_u+\e)}=
e^{-\e\partial_\e} \frac 1{(K+\phi_u+\e)}
\label{infepsider}
\ee
where $e^{-\e\partial_\e}$ means
\be
e^{-\e\partial_\e}= e^{- a\partial_\e}\vert_{a=\e}\0
\ee

This expansion\footnote{The difference between the $\e$ and $\ve$ parameters is as follows: $\ve$ is a regulator we use in order to define the solution $\psi_u^{\ve}$ in the limit $\ve\to 0$; $\e$ is a pure mathematical expansion parameter that helps us monitoring the consistency of the formalism.} has the advantage that it expresses $\frac 1{K+\phi_u}$ in terms of $\frac 1{(K+\phi_u+\e)}$. The latter, as was shown in \cite{BGT1}, does not suffer from the same (would-be) ambiguity
as $\frac 1{K+\phi_u}$. We can write in general
\be
\frac 1{K+\phi_u+\e}= \int_0^\infty dt\, e^{-t(K+\phi_u+\e)}\label{Schrightepsilon}
\ee
So we will use (\ref{infepsisum}) or (\ref{infepsider}) as our definition of $\frac 1{K+\phi_u}$.
Of course now we have to pay attention that the series in (\ref{infepsisum}) and (\ref{infepsider}) converge, or that
the shift operator $e^{-\e\partial_\e}$ acts on objects whose dependence on $\e$ is regular. It should be noticed that (\ref{infepsider}) may be interpreted as 
\be
{\rm either} \quad\quad e^{-\e\partial_\e} \int_0^\infty dt\, e^{-t(K+\phi_u+\e)}, \quad\quad \quad{\rm or}
\quad \quad\int_0^\infty dt\,e^{-\e\partial_\e}  e^{-t(K+\phi_u+\e)}\label{ambig}
\ee
These are the same in case of regularity, but may give rise to an ambiguity otherwise. The just mentioned ambiguities will be the main focus of this section.

Let us see an example straightaway. According to this new representation the lump solution can be written as
\begin{align}\label{expan}
\phi_u\to\psi_{u,\e}=
c\phi_u- \sum_{n=0}^\infty \frac {\e^n}{(K+\phi_u+\e)^{n+1}}(\phi_u-\delta\phi_u)Bc\partial c
\end{align}
One of the basic expressions considered in the introduction is $\frac 1{K+\phi_u}(\phi_u-\delta\phi_u)$,
which must be nonsingular. Let us check it on the basis of the new representation
\begin{align}\label{firstcheck1}
{\rm Tr} \left[\frac 1{K+\phi_u}(\phi_u-\delta\phi_u)\right]
&=\sum_{n=0}^\infty{\rm
Tr}\big[\frac{\ve^{n}}{(K+\phi_u+\e)^{n+1}}(\phi_u-\delta\phi_u)\big]\0\\
&=\sum_{n=0}^\infty\frac{(-\e)^{n}}{n!}\,\partial_\e^n{\rm
Tr}\big[\frac1{K+\phi_u+\e}(\phi_u-\delta\phi_u)\big]\0\\
 &=\sum_{n=0}^\infty\frac{(-\e)^{n}}{n!}\,\partial_\e^n\int_0^{\infty}dt~e^{-\e t}{\rm
Tr}\big[u\partial_u\phi_ue^{-t(K+\phi_u)}\big]\0\\
&=-\sum_{n=0}^\infty\frac{(-\e)^{n}}{n!}\,\partial_\e^n
\int_0^{\infty}dt~e^{-\e t}\frac u{t}\partial_u{\rm
Tr}\big[e^{-t(K+\phi_u)}\big]\0\\
&=- e^{-\e\partial_\e}\, \int_0^{\infty}dt~e^{-\e
t}\frac u{t}\partial_ug(ut)
\end{align}
where $g(u)$ has been defined above. Setting $y=ut$ we can write
\begin{align}\label{firstcheck2}
{\rm Tr} \left[\frac 1{K+\phi}(\phi-\delta\phi_u)\right] &
=-e^{-\e\partial_\e} \int_0^{\infty}dy~e^{-\e
y/u}\,\partial_yg(y)\0\\
&=-\int_0^{\infty}dy~e^{-\e\partial_\e} e^{-\e y/u} \partial_yg(y)=-\int_0^{\infty}dy\partial_yg(y)
\end{align}
This is the result we would obtain by using directly (\ref{Schright}).

Two remarks are in order.
First, in the passage from the first to the second line of (\ref{firstcheck2}) we exchange integration and summation: this is allowed if the integral is convergent without the $e^{-\e y/u}$ factor\footnote{Actually the integral could be less than convergent, even be logarithmically divergent in the IR, but in the sequel we will not meet such an occurrence.}. The integrand we are considering, $ \partial_yg(y)$, behaves like $1/y^2$ for large $y$, so this condition is satisfied in the IR ($s\to \infty$). Alternatively one can analyse the applicability of the shift operator $e^{-\e\partial_\e}$ to the integral $\int_0^{\infty}dy~e^{-\e
y/u}\,\partial_yg(y)$. This is correct as long as the integral is differentiable as a function of $\e$,
which is the case when $\int_0^{\infty}dy~ \,\partial_yg(y)$ is convergent.

As for the UV, $y\approx 0$, (and this is the second remark) we know that a subtraction is needed, but it involves only the 0-th order term of the summation and it does not depend on $\e$. The 0-th order term can be easily treated separately and the relevant subtraction is precisely the same as the one needed in the RHS of the second line of (\ref{firstcheck2}).

A useful remark, which has already been exploited in section 2, is the following one: the argument of ${\rm Tr}$ in eqs.(\ref{firstcheck1},\ref{firstcheck2}) is the essential matter ingredient of $\psi_u$. The result
(\ref{firstcheck2}) means that, when inserted into a correlator, the solution $\psi_u$ leads to a contribution
$\sim \frac 1{\sqrt y}$ in the UV. This has been confirmed by all the results in \cite{BGT1,BGT2} and in the rest of this paper, and holds for $\psi_u^{\ve}$ as well, as it is easy to check.

The convergence of the integral in the IR in the previous example is crucial. As a counterexample let us consider
\begin{align}\label{secondcheck}
{\rm Tr} \left[\frac 1{K+\phi_u} \right]
&=\sum_{n=0}^\infty{\rm
Tr}\big[\frac{\e^{n}}{(K+\phi_u+\e)^{n+1}}\big]
=\sum_{n=0}^\infty\frac{(-\e)^{n}}{n!}\,\partial_\e^n{\rm
Tr}\big[\frac1{K+\phi_u+\e} \big]\0\\
 &=\sum_{n=0}^\infty\frac{(-\e)^{n}}{n!}\,\partial_\e^n\int_0^{\infty}dt~e^{-\e t}{\rm
Tr}\big[ e^{-t(K+\phi_u)}\big]\0\\
&=-\sum_{n=0}^\infty\frac{(-\e)^{n}}{n!}\,\partial_\e^n\int_0^{\infty}dt~e^{-\e
t}\,g(ut)
\end{align}
The integral diverges because $g(t)\to 1$ as $t\to \infty$, as a consequence we cannot exchange summation
and integration. It is easy to show that the summation diverges even in the presence of the $e^{-\e
t}$ in the integrand. Alternatively one can argue that whether we apply $e^{-\e \frac {\partial}{\partial\e}}$ 
inside or outside the integral, the result is infinite.

The new representation agrees with the old one (see \cite{BMT}) on the fact that $\frac 1{K+\phi_u}$ is singular.
 
\subsection{The energy}

Of course it is very important that with the new representation we are able to show that we obtain
for the energy the same result as in \cite{BGT1}. We recall that the energy expression for $\psi_u$
was obtained by means of the replacement (\ref{Efinal}) and takes the following form
\be
E[\psi_u] = -\frac 16 \langle\psi_u\psi_u\psi_u\rangle= \int_0^\infty ds \, F(s)\label{Epsiu}
\ee
where $F(s)$ behaves like $\sim 1/s^2$ for large $s$ and needs a subtraction in the UV. With the new
representation we get
\begin{align}
E[\psi_u]=&-\frac 16 \langle\psi_u\psi_u\psi_u\rangle=\big(
\sum_{n=0}^\infty(-1)^n\frac{\e^n}{n!}\,\partial^n_\e\big){\rm
Tr}\Big[\Big(\frac1{K+\phi_u+\e}(\phi_u-\delta\phi_u)Bc\partial
c\Big)^3\Big]\0\\
=&
\sum_{n=0}^\infty(-1)^n\frac{\e^n}{n!}\partial^n_\e\int_0^\infty
ds~ e^{-\e s/2u}F(s)=\int_0^\infty ds~ e^{-\e
s/2u}\sum_{n=0}^\infty\frac1{n!}\Big(\frac{\e
s}{2u}\Big)^nF(s)\0\\
=&\int_0^\infty dsF(s).\label{Epsiu1}
\end{align}
that is, the same result as (\ref{Epsiu}).

In the previous derivation there are a few nontrivial passages. The first is the passage from the second
to the third member, which however involves standard manipulations. Another nontrivial step is
the one in the second line (i.e. exchanging summation and integration). This is possible only because
the integrand is convergent in the IR and the singularity near $s\approx 0$ is present only in the zeroth order term. The latter can be easily dealt with separately from the others, it can be isolated and subtracted, and this subtraction is independent of $\e$ and is the same mentioned above for $F(s)$.

In conclusion the new representation gives for $E[\psi_u]$ the same result obtained in \cite{BGT1} by using
the Schwinger representation (\ref{Schright}) for $\frac1{K+\phi_u}$.

In order to complete our derivation we have to consider the last two terms in (\ref{TVlumpenergy})
(beware the different roles of $\ve$ and $\e$)
\be
&& -3\langle\psi_u\psi_u\psi_u^{\ve}\rangle\0\\
&&~~~~~=3\sum_{n=0}^\infty\frac{(-\e)^n}{n!}\partial^n_\e{\rm
Tr}\Big[\Big(\frac1{K+\phi_u+\e}(\phi_u-\delta\phi_u)Bc\partial
c\Big)^2\frac1{K+\phi_u+\ve}(\phi_u+\ve-\delta\phi_u)Bc\partial
c\Big]\0\\
&&~~~~~=3\sum_{n=0}^\infty\frac{(-\e)^n}{n!}\partial^n_\e\int
ds\,dx \,dy\, e^{-\e s\frac{x+y}{2u}}F_1(\ve,s,x,y)\0\\
&&~~~~~=3\int ds\,dx \,dy\, e^{-\e
s\frac{x+y}{2u}}\sum_{n=0}^\infty\frac1{n!}
\big(s\frac{x+y}{2u}\big)^nF_1(\ve,s,x,y)\0\\
&&~~~~~=3\int ds\,dx \,dy\, F_1(\ve,s,x,y)\label{uue}
\ee
where $F_1(\ve,s,x,y)$ is related to the integrand of the functional
$\langle\psi_u\psi_u\psi_u^{\ve}\rangle$ calculated in \cite{BGT1}
by using Schwinger representation (\ref{Schright}) for $\frac1{K+\phi_u}$. We
have introduced the usual variables
\begin{align}
T=t_1+t_2+t_3,\quad x=\frac{t_1}T,\quad y=\frac{t_2}T,\quad s=2uT
\end{align}
and one can identify $F_1(\ve,s,x,y)$  with the integrand of eq.(9.5) in \cite{BGT1} after
the change of variables $x\leftrightarrow y$ followed by $y\rightarrow 1-y$.

Similarly

\be
&& 3\langle\psi_u\psi_u^{\ve}\psi_u^{\ve}\rangle\0\\
&&~~~~~=
-3\sum_{n=0}^\infty\frac{(-\e)^n}{n!}\partial^n_\e) {\rm
Tr}\Big[\frac1{K+\phi_u+\e}(\phi_u-\delta\phi_u)Bc\partial
c\Big(\frac1{K+\phi_u+\ve}(\phi_u+\ve-\delta\phi_u)Bc\partial
c\Big)^2\Big]\0\\
&&~~~~~=-3\sum_{n=0}^\infty\frac{(-\e)^n}{n!}\partial^n_\e\int
ds\,dx \,dy\, e^{-\e s\frac{x}{2u}}F_2(\ve,s,x,y)\0\\
&&~~~~~=-3\int ds\,dx \,dy\, e^{-\e
s\frac{x}{2u}}\sum_{n=0}^\infty\frac1{n!}
\big(s\frac{x}{2u}\big)^nF_2(\ve,s,x,y)\0\\
&&~~~~~=-3\int ds\,dx \,dy\, F_2(\ve,s,x,y)\label{uee}
\ee
Once again $F_2(\ve,s,x,y)$ is related to the integrand of the functional
$\langle\psi_u\psi_u^{\ve}\psi_u^{\ve}\rangle$ calculated in \cite{BGT1}
by using Schwinger representation (\ref{Schright}) for
$\frac1{K+\phi_u}$. Identification of $F_1(\ve,s,x,y)$  with the integrand of eq.(9.6) in \cite{BGT1}
requires the same change of variables as above.

As for the UV, the same holds as for (\ref{Epsiu}): the subtraction is $\e$-independent, it involves only 0-th order terms in each series and can be treated separately.

Summarizing, the results obtained with the new representation (\ref{infepsisum},\ref{infepsider}) for eq.(\ref{TVlumpenergy}) are the same as the results we obtained by straightforwardly using the Schwinger representation (\ref{Schright}) for $\frac1{K+\phi_u}$ in \cite{BGT1}.

\subsection{About the closed string overlap}

In \cite{BMT} it was shown that the $\psi_u$ solution can satisfy the closed string overlap condition.
Now we are in a position to clarify some aspects of this problem. The closed string overlap (CSO) is closely related to the
traces we have considered above. Since
the contribution from the identity piece of the solution is zero the
CSO is given by

\begin{align}
{\rm Tr}[V_c\psi_u]=&-\sum_{n=0}^\infty{\rm
Tr}\big[V_c\frac{\e^{n}}{(K+\phi_u+\e)^{n+1}}
(\phi_u-\delta\phi_u)Bc\partial c\big]\0\\
&=-\sum_{n=0}^\infty\frac{(-\e)^{n}}{n!}\partial_\e^n{\rm
Tr}\big[V_c\frac1{K+\phi_u+\e}(\phi_u-\delta\phi_u)Bc\partial c\big]\0\\
&=-\sum_{n=0}^\infty\frac{(-\e)^{n}}{n!}\partial_\e^n
\int_0^{\infty}dt~e^{-\e t}{\rm
Tr}\big[V_ce^{-t(K+\phi_u)}(\phi_u-\delta\phi_u)Bc\partial c\big]\0\\
&=-\int_0^{\infty}dt~e^{-\e t}\sum_{n=0}^\infty\frac{(\e
t)^{n}}{n!}{\rm
Tr}\big[V_ce^{-t(K+\phi_u)}(\phi_u-\delta\phi_u)Bc\partial c\big]\0\\
&=- e^{-\e \partial_\e}
\int_0^{\infty}dt~e^{-\e t}{\rm
Tr}\big[V_ce^{-t(K+\phi_u)}(\phi_u-\delta\phi_u)Bc\partial c\big]\0\\
&=-\int_0^{\infty}dt~{\rm
Tr}\big[V_ce^{-t(K+\phi_u)}(\phi_u-\delta\phi_u)Bc\partial c\big]\0\\
&=\langle\langle V_c\rangle\rangle^{UV}- \langle\langle V_c\rangle\rangle^{IR}\label{CSO}
\end{align}
The last step follows from \cite{BMT}. This is the result we would obtain by using the Schwinger
representation directly for $\frac 1{K+\phi_u}$, as was done in \cite{BMT}. Once again all this is correct if, in the above expressions, the integrand, when $e^{-\e t}$ is replaced by 1, is convergent, i.e. if the
integral in the last line of (\ref{CSO}) is continuous also at $\e=0$. Obviously this depends on $V_c$. It implies in particular that $\langle\langle V_c\rangle\rangle^{IR}$ is finite. Anyhow, in this problem
one should take into account the very likely presence of a UV divergence and the relative subtraction. We believe
the CSO question should be reconsidered in the light of concrete cases, as we have done above and in \cite{BGT1} with the energy of the lump.

\section{Concerning the identity $\frac 1{K+\phi_u}(K+\phi_u)=I$}

Let us return to section 2 and  eqs.(\ref{IDDu}), (\ref{Schright}) and  (\ref{Schwrong}). Applying our new representation we get
\be
\frac1{K+\phi_u}(K+\phi_u)&=& \sum_{n=0}^\infty \frac {(-\e)^n}{n!}\,\partial_\e^n \frac 1{K+\phi_u+\e}(K+\phi_u)\0\\
&=& e^{-\e \partial_\e}\left( 1 - \frac {\e}{K+\phi_u+\e}\right)\0\\
&=& 1- e^{-\e \partial_\e} \frac {\e}{K+\phi_u+\e}\label{IDDu1}
\ee
The expression $e^{-\e \partial_\e} \frac {\e}{K+\phi_u+\e}= \lim_{\e\to 0} \frac {\e}{K+\phi_u+\e}$ is a more appropriate way to write $\Omega_u^\infty$ (it is extremely helpful to keep in mind the analogy with $\Omega(r)$ in sec. 3). It is of course formally vanishing, but  to make any sense of such  an expression one has to evaluate it in correlators. For instance, taking the trace, as in section 2, we are led to evaluate
\be
{\rm Tr} \left[ \frac {\e}{K+\phi_u+\e}\right]= \e \int_0^\infty  dt\, e^{-\e t}\, g(ut)\label{IDDu2}
\ee
Since, once again, $g(\infty)=1$, the limit $\e\to 0$ is not continuous, and this depends on the fact that, as we have seen many times, the integral in the RHS of (\ref{IDDu2}) is (linearly) divergent when the factor $e^{-\e t}$ is replaced by 1. As a consequence the shift
operator $e^{-\e \partial_\e} $ cannot be applied in a consistent way in (\ref{IDDu2}). In fact it is not clear
what value one should assign to the expression
\be 
e^{-\e \partial_\e}\left(\e \int_0^\infty  dt\, e^{-\e t}\, g(ut)\right)
\ee
depending on whether we integrate first or apply first the operation $ e^{-\e \partial_\e} \e$ to the integrand.

On the other hand, if (\ref{IDDu1}) is inserted in a correlator (like the energy one) where the integrand
without the exponential factor decreases fast enough, then the result of the application of $e^{-\e \partial_\e} $ to $ \frac {\e}{K+\phi_u+\e}$ is unambiguously 0. 
This can be seen by considering for instance the following contraction  
\begin{align}
 &{\rm Tr}\Big[\d^2c~e^{-(K+\phi)}e^{-\e
\partial_\e} \left(\frac {\e}{(K+\phi_u+\e)} (\phi_u-\delta\phi_u)
c\partial c\right)\Big]\label{RHS1}\\
&=e^{-\e
\partial_\e}\e\int_0^\infty dt~e^{-\e t}{\rm
Tr}\big[(\phi_u-\delta\phi_u)e^{-(t+1)(K+\phi)}\big]
\langle\d^2c(t+1)c\d c(0)\rangle_{C_{t+1}}\0\\
&=e^{-\e
\partial_\e}\e\int_0^\infty dt~e^{-\e
t}\langle(\phi_u(0)-\delta\phi_u(0))e^{-\int_0^{t+1}ds\phi(s)}\rangle_{C_{t+1}}
\langle\d^2c(t+1)c\d c(0)\rangle_{C_{t+1}}\0\\
 &=- e^{-\e\partial_\e}\e\int_0^{\infty}dt~e^{-\e t}G(t)\frac
u{t+1}\partial_ug\big(u(t+1)\big)=
2e^{-\e\partial_\e}\e\int_0^{\infty}dt~e^{-\e t}\frac
u{t+1}\partial_ug\big(u(t+1)\big)\0
 \end{align}
 where the ghost contribution is given by
\be 
G(t)= \langle \d^2c(t+1) (c\d c)(0)\rangle_{C_{t+1}}=-2.\0 
\ee
Now we can write eq.(\ref{RHS1}) as
 \be
&&2\big(e^{-\e\partial_\e}\e\big)e^{-\e\partial_\e}\int_0^{\infty}dt~e^{-\ve t}\frac
u{t+1}\partial_ug\big(u(t+1)\big)\0\\
&&~~~~~~~~~~~~~~~~~~~~~~~~~=2\big(e^{-\e\partial_\e}\e\big)\int_0^{\infty}dt~\frac
u{t+1}\partial_ug\big(u(t+1)\big)=0.\label{RHS2}
\ee 
We note that this last result does not need any UV subtraction.

\subsection{How to compute correlators with spurious terms}

After these long preliminaries let us come to the would-be violation of the equation of motion due to the second term in the RHS of (\ref{IDDu1}), pointed out in \cite{EM}. To this end we rewrite
\be
\psi_u \rightarrow \psi_{u,\e}&=c\phi_u-  e^{-\e \partial_\e}\frac 1{(K+\phi_u+\e)}(\phi_u-\delta\phi_u)Bc\partial c\label{psiue}
\ee
and apply $Q$ to it. Using in particular
\be
Q \left(e^{-\e \partial_\e}\frac 1{(K+\phi_u+\e)}\right) = - e^{-\e \partial_\e} \frac 1{(K+\phi_u+\e)} (Q\phi_u)
\frac 1{(K+\phi_u+\e)}\label{Qpsiue}
\ee
and proceeding as in section 3.2 of \cite{BMT}, we find
\be
&&Q \psi_{u,\e}=Q\left(   c\phi_u-  e^{-\e \partial_\e}\frac 1{(K+\phi_u+\e)}(\phi_u-\delta\phi_u)Bc\partial c   \right)\label{EOMe}\\
&& =e^{-\e \partial_\e} \left[ 1 + \frac 1{(K+\phi_u+\e)} (c\partial \phi_u +\partial c \delta\phi_u)
\frac 1{(K+\phi_u+\e)} B -
\frac 1{(K+\phi_u+\e)} K \right] (\phi_u-\delta\phi_u) c\partial c\0\\
&&=e^{-\e \partial_\e}\left[ \left( c \phi_u -\frac 1{(K+\phi_u+\e)} (\phi_u-\delta\phi_u) \partial c\right)
\frac 1{(K+\phi_u+\e)}+
\frac {\e}{(K+\phi_u+\e)}c  \right](\phi_u-\delta\phi_u) B c\partial c\0\\
&&= -  \psi_{u,\e}\psi_{u,\e}  + e^{-\e \partial_\e} \left(\frac {\e}{(K+\phi_u+\e)} (\phi_u-\delta\phi_u) c\partial c\right)\0
\ee
In a regular setting, that is when inserted in a correlator regular in $\e$, this boils down to the usual
eom $Q\psi_u = -\psi_u\psi_u$, and in particular the second piece in the RHS of the last line vanishes. Let us see what happens if we, nevertheless, insist in keeping (\ref{EOMe}) in the expression of the energy. We have
\be
-\langle \psi_u Q\psi_u\rangle &\rightarrow& -\langle \psi_{u,\e} Q\psi_{u,\e}\rangle \label{Qenergy}\\
&=& \langle \psi_{u,\e} \psi_{u,\e}\psi_{u,\e}\rangle+ \langle
\psi_{u,\e}  e^{-\e \partial_\e}\left(\frac {\e}{(K+\phi_u+\e)}
(\phi_u-\delta\phi_u) c\partial c\right) \rangle\0 
\ee 
The second term in the RHS equals 
\be 
e^{-\e \partial_\e}\langle\frac
1{(K+\phi_u+\e)}(\phi_u-\delta\phi_u)Bc\partial c\,\, \frac
{\e}{(K+\phi_u+\e)} (\phi_u-\delta\phi_u) c\partial c\rangle\label{second}
\ee
With the usual procedure we can write this as ($T=t_1+t_2$)
\be 
e^{-\e \partial_\e}\left(\e\int_{0}^\infty dt_1dt_2~e^{-\e T}{\cal G}(t_1,t_2)u^2g(uT)
\Big\{\Big(-\frac{\partial_{uT}g(uT)}{g(uT)}\Big)^2
+2G_{2uT}^2(\frac{2\pi t_1}T)\Big\}\right),\label{Qenergy1}
\ee
where the ghost part is given by 
\be 
{\cal G}(t_1,t_2)= \langle (B c\partial c)(t_1)
(c\partial c)(0)\rangle_{C_T} =\frac{t_1}{\pi}\sin(\frac{2\pi t_1}T)
-\frac{2T}{\pi^2}\sin^2(\frac{\pi t_1}T).\label{Qenergy2}
\ee 
Let us show now that (\ref{Qenergy1}) reduces to the form
\be 
e^{-\e \partial_\e}\left(\e\int_{0}^\infty ds\, e^{-\e s} \EF(s)\right)\label{Qenergy3}
\ee
where $ \EF(s)\to {\rm const}$ for large $s$ and the integral is UV finite.

Denoting $x=\frac {t_1}T$, Eq.(\ref{Qenergy2}) can be rewritten as 
\be \label{B1}
e^{-\vareta\partial_{\vareta}}\vareta\int_{0}^\infty ds s^2 \int_{0}^1 dx{\cal E}(x)~e^{-\vareta s}\g(s)
\Big\{\Big(-\frac{\partial_{s}\g(s)}{\g(s)}\Big)^2
+\frac12G_{s}^2(2\pi x)\Big\},
\ee
where $\vareta=\frac {\e}{2u}$ and 
\be
{\cal E}(x)=\langle (B c\partial c)(x)(c\partial c)(0)\rangle_{C_1} =\frac{-1+\text{cos}(2 \pi  x)+\pi  x \text{sin}(2 \pi  x)}{\pi ^2}\label{B2}
\ee
Since $\int_{0}^1 dx{\cal E}(x)=-\frac{3}{2 \pi ^2}$, the term with no $G_s$ is given by
\be
-\frac{3}{2 \pi ^2}\vareta\int_{0}^\infty ds s^2~e^{-\vareta s}\g(s)\Big(-\frac{\partial_{s}\g(s)}{\g(s)}\Big)^2\label{B3}
\ee
As $\g(s)\approx\frac{1}{\sqrt{s}}$ in the UV we are in the case of eq.(8.13) of \cite{BGT1}  and so the UV contribution vanishes for $\vareta \to 0$. In the IR we are in the case of eq.(8.17) of \cite{BGT1} and so the IR contribution vanishes too. It can be easily proven that 
\be 
3\int_0^1dx{\cal E}(x)G_s^2(2\pi x)&=& \frac4{\pi}\int_0^1dy\int_0^ydx\, \sin\pi x\,\sin\pi y\,\sin\pi(x-y)\0\\
&&\quad\quad\cdot\Big(G_{s}^2(2\pi
x)+G_{s}^2(2\pi(x-y))+G_{s}^2(2\pi y)\Big)\label{GU2}
\ee 
where the expression in the RHS is the same  as eq.(3.7) of \cite{BGT1}. Therefore we have
\be 
&&e^{-\vareta
\partial_{\vareta}}\left(\frac12\vareta\int_{0}^\infty ds s^2 ~e^{-\vareta s}\g(s)\int_{0}^1 dx{\cal E}(x,0)G_{s}^2(2\pi x)\right)\0\\
&=&e^{-\vareta
\partial_{\vareta}}\left(\frac16\vareta\int_{0}^\infty ds s^2 ~e^{-\vareta s}\g(s)\frac4\pi\int_0^1dy\int_0^ydx\, \sin\pi x\,\sin\pi y\,\sin\pi(x-y)\right.\0\\
&&\quad\quad\cdot\left.\Big(G_{s}^2(2\pi
x)+G_{s}^2(2\pi(x-y))+G_{s}^2(2\pi y)\Big)\right)\label{GU3} 
\ee
We can now avail ourselves of the results in \cite{BGT1}. The integration over $x$ and $y$ leads to an integrand in $s$ that behaves like a constant for large $s$, if one abstracts from the
factor $e^{-\vareta s}$. Thus we have obtained (\ref{Qenergy3}). Under these conditions the limit for $\vareta\to 0$ of the $s$ integral is discontinuous and we are not allowed to exchange $e^{-\vareta \partial_{\vareta}}$ with the integration. We do not know what value should be assigned to (\ref{Qenergy3}).  As a consequence
the additional piece in RHS of (\ref{Qenergy}) cannot be assigned an unambiguous value without an {\it ad hoc} prescription.

With an {\it ad hoc} prescription we can still obtain 
a finite result. If, for instance, we first multiply $\vareta$ by the result of the integration and subsequently apply $e^{-\vareta \partial_{\vareta}}$ we obtain $-2\beta$, where $\beta$ is the number introduced in \cite{BGT1}, see also sec.2 above. This result is the same as the one obtained by \cite{EM}. {\it But one should not forget that it is prescription-dependent}. We remark, in addition, that the term (\ref{GU3}) appears in the RHS of eq.(\ref{Qenergy}) together with $\langle \psi_{u,\e} \psi_{u,\e}\psi_{u,\e}\rangle$. The latter is a UV divergent term (in \cite{BGT2} it is even gauge-dependent) and needs a UV subtraction, which, we recall, carries some arbitrariness into the problem. For instance, one could choose the UV subtraction in such a way as to kill the contribution of $-2\beta$ altogether and there would be no violation of the EOM. Therefore it is not even clear what the would-be violation of the EOM means.

The ambiguity intrinsic in this problem reminds us of the discussion after eq.(\ref{wrong1}) in sec. 3.
There, by integrating a vanishing distribution over a non test function, we could obtain a nonvanishing result. This is no accident. The nonvanishing of the second term in the RHS of (\ref{Qenergy}) is analogous. The
string field  $e^{-\e \partial_\e} \frac {\e}{K+\phi_u+\e}(\phi_u-\delta \phi_u)$ plays the role of the
vanishing distribution and $\psi_{u,\e}$ the role of the singular test function. The only difference here is
that the singularity comes from the IR, because of the inversion of roles introduced by the Schwinger
representation. In this regard we can be more precise. If we strip (\ref{second}) of the $\e$ factor in the numerator, what remains represents the string field $\frac 1 {K+\phi_u+\e}(\phi_u-\delta \phi_u)$ contracted with itself,
which can be interpreted as the `norm' square of this string field, in the limit $\e\to 0$. Well, the above results tell us that this `norm' is infinite. It is this infinity that multiplied by the stripped factor $\e$
allows us to obtain the above finite result. This clearly confirms the singular nature of $\psi_{u,\e}$  as a test state.

It is instead possible to derive a prescription-independent (and subtraction-independent) result, even taking into account the spurious term, provided one proceeds in another way. Let us rewrite $\Phi_0^\ve$, eq. (\ref{psiupsie}), using the new representation: $\Phi(\e,\ve)=\psi_{u,\e}-\psi_u^\ve$, where $\psi_u^\ve$, in the $\ve\to 0$ limit, is the tachyon vacuum solution defined in \cite{BGT1}. We get
\be
Q \psi_{u,\e}&=&-  \psi_{u,\e}\psi_{u,\e}  + e^{-\e \partial_\e} \left(\frac {\e}{(K+\phi_u+\e)} (\phi_u-\delta\phi_u) c\partial c\right)\0\\
Q \psi_u^\ve &=& -\psi_u^\ve\psi_u^\ve\label{B4}\\
{\cal Q} \Phi(\e,\ve)&=&- \Phi(\e,\ve)\Phi(\e,\ve) + e^{-\e \partial_\e} \left(\frac {\e}{(K+\phi_u+\e)} (\phi_u-\delta\phi_u) c\partial c\right)\0
\ee
where ${\cal Q}\Phi=Q\Phi+\psi_u^\ve\Phi+\Phi\psi_u^\ve $. Moreover
\be
-\langle  \Phi(\e,\ve) {\cal Q} \Phi(\e,\ve)\rangle
&=& \langle  \Phi(\e,\ve)  \Phi(\e,\ve) \Phi(\e,\ve)\rangle\label{B5}\\
&&+ \langle
 \Phi(\e,\ve)  e^{-\e \partial_\e}\left(\frac {\e}{(K+\phi_u+\e)}
(\phi_u-\delta\phi_u) c\partial c\right) \rangle\0 
\ee 
If we use the just defined representation, the second term in the RHS equals 
\be 
&&e^{-\e \partial_\e}  \langle \psi_{u,\e} \left(\frac {\e}{(K+\phi_u+\e)}
(\phi_u-\delta\phi_u) c\partial c\right)\rangle - e^{-\e \partial_\e} \langle \psi_u^\ve \left(\frac {\e}{(K+\phi_u+\e)}
(\phi_u-\delta\phi_u) c\partial c\right)\rangle\0\\
&=&-2\beta -e^{-\e \partial_\e}\langle\frac
1{(K+\phi_u+\ve)}(\phi_u+\ve-\delta\phi_u)Bc\partial c\,\, \frac
{\e}{(K+\phi_u+\e)} (\phi_u-\delta\phi_u) c\partial c\rangle\0 \\
&=&-2\beta -e^{-\e \partial_\e}\langle\frac
1{(K+\phi_u+\ve)}(\phi_u-\delta\phi_u)Bc\partial c\,\, \frac
{\e}{(K+\phi_u+\e)} (\phi_u-\delta\phi_u) c\partial c\rangle\0\\
&& -e^{-\e \partial_\e}\langle\frac
\ve{(K+\phi_u+\ve)}Bc\partial c\,\, \frac
{\e}{(K+\phi_u+\e)} (\phi_u-\delta\phi_u) c\partial c\rangle\label{B6}
\ee
In (\ref{B6}) there is no need of UV subtractions. 
The last two terms in the RHS equal, respectively,
\be
&&e^{-\e \partial_\e}\langle\frac
1{(K+\phi_u+\ve)}(\phi_u-\delta\phi_u)Bc\partial c\,\, \frac
{\e}{(K+\phi_u+\e)} (\phi_u-\delta\phi_u) c\partial c\rangle\label{B7}\\
&=&e^{-\e
\partial_\e}\left(\e\int_{0}^\infty dt_1dt_2~e^{-\ve t_2-\e t_1}{\cal G}(t_1,t_2)u^2g(uT)
\Big\{\Big(-\frac{\partial_{uT}g(uT)}{g(uT)}\Big)^2
+2G_{2uT}^2(\frac{2\pi t_1}T)\Big\}\right)\0\\
&=&e^{-\vareta
\partial_{\vareta}}\left(\vareta\int_{0}^\infty ds s^2 \int_{0}^1 dx\,{\cal E}(x)~e^{-T(\ve(1-x)+\e x)}\g(s)
\Big\{\Big(-\frac{\partial_{s}\g(s)}{\g(s)}\Big)^2
+\frac12G_{s}^2(2\pi x)\Big\}\right)\0\\
&=&e^{-\vareta
\partial_{\vareta}}\left(\vareta\int_{0}^\infty ds s^2 \int_{0}^1 dx\,~e^{-\vareta s}{\cal E}(1-x)
~e^{s\frac{\e-\ve}{2u} x}\g(s)
\Big\{\Big(-\frac{\partial_{s}\g(s)}{\g(s)}\Big)^2
+\frac12G_{s}^2(2\pi x)\Big\}\right)\0
\ee
and
\be
&& e^{-\e \partial_\e}\langle\frac
\ve{(K+\phi_u+\ve)}Bc\partial c\,\, \frac
{\e}{(K+\phi_u+\e)} (\phi_u-\delta\phi_u) c\partial c\rangle\label{B8}\\
&=&e^{-\e \partial_\e}\left(\ve\e\int_0^\infty dt_1dt_2~e^{-\ve t_2-\e t_1}{\cal G}(t_1,t_2)\frac{u}{t_1+t_2}\partial_u g(uT)\right)\0\\
&=&e^{-\e \partial_\e}\left(\ve\e\int_0^\infty dT\,T\int_0^1dx~e^{-T(\ve(1-x)+\e x)}
{\cal E}(x)u\partial_u g(uT)\right)\0\\
&=&e^{-\vareta
\partial_{\vareta}}\left(\vareta\frac{\ve}{2u}\int_0^\infty ds s^2~e^{-\vareta s} \int_{0}^1dx~
{\cal E}(1-x)~e^{s\frac{\e-\ve}{2u} x}\partial_s\g(s)\right)\0
\ee
As we have learnt in section 2 these quantities must be evaluated in the limit $\ve\to 0$. We are by now very familiar with this type of integrals and can easily come to the conclusion that both angular integrations are finite even without the $e^{s\frac{\e-\ve}{2u} x}$ factors so that in the limit $\ve,\e\to 0$ the integration is continuous in $\ve,\e$ and such factors can be dropped. Thus, using always the same representation, the former integral is just $-2\beta$. The latter is the same as eq.(4.21) of \cite{EM}. It is convergent both in the UV and the IR.

So we find
\be 
\lim_{\e\to 0}\langle \Phi(\e,\ve)  e^{-\e \partial_\e}\left(\frac {\e}{(K+\phi_u+\e)}
(\phi_u-\delta\phi_u) c\partial c\right) \rangle =-2\beta+2\beta-0=0\0
\ee
This is a prescription-independent  (and subtraction-independent) result, the reason being that the 
overall $s$ integrand has,  in the limit $\ve\to 0$, the right convergent behaviour for large $s$ in order to guarantee continuity in $\e$ also at $\e=0$\footnote{What happens here is that we have the difference of two integrals which are divergent (without the $e^{-\vareta s}$) but the divergences cancel each other in the limit
$\ve\to 0$.}.
We deduce that {\it this is the right way to compute the lump energy}, {\it and}, as by now should be obvious, {\it the spurious term does not contribute to it}.

The term $\Gamma(\e)\equiv e^{-\e \partial_\e} \frac {\e}{K+\phi_u+\e}(\phi_u-\delta \phi_u)c\del c $ in the RHS of (\ref{B5}) is clearly similar to the spurious terms considered
in connection with the solution to the Einstein EOM in section 3. It violates the principle of continuity and (as a natural consequence) it is ambiguous. However when, in spite of this, it is taken into account in our calculation of the energy outlined in the introduction, it yields a (non-ambiguous) vanishing contribution, as we have just shown, because the integral it is inserted in is convergent (even without the $e^{-\vareta s}$ factor). When inserted into non-convergent integrals, in the limit $\ve\to 0$ it gives rise to an ambiguous term, see (\ref{Qenergy3}) above. 
In a well-defined setting, provided by distribution theory, the nature of this term is clear: it is a spurious term and {\it should not} be taken into account. In the language of distribution theory, $\phi_u^{\ve}$ and $\psi_{u,\e}$ are not good test states because of their asymptotic behaviors, but their difference is.

\section{Good test string fields}

So far we have seen few example of good test states: one is the state defined implicitly by eq.(\ref{RHS1}), another is in the second line of eq.(\ref{B5}) and others, possibly, in the discussion of the CSO.
A question one might ask is whether there are enough good test states in the theory. This 
is connected with the problem of Fock space states. 
It is customary in SFT to verify a string field's properties by contracting it
with Fock space states, the latter being considered a large enough set of states (a completeness). The question of whether $\Gamma(\e)=e^{-\e \partial_\e} \frac {\e}{K+\phi_u+\e}(\phi_u-\delta\phi_u)c\del c$ when contracted with a large enough set of states vanishes must be formulated in the appropriate way. These states cannot be 
`naked' Fock space (see some examples of them in \cite{EM}) because such states {\it are not good test states}.
Once again it is worth recalling that if we contract a formally vanishing distribution with a non-test state 
we can get something nonvanishing. 
First of all the states we are looking for must be such that the resulting contractions with $\Gamma(\e)$ be nonsingular (with respect to singularities due to collapsing points). But, especially, they must be characterized
by integrable behaviour in the UV and, ignoring the overall $e^{-\e t}$ factor, in the IR. It is in fact self-evident that {\it all} the states with such properties annihilate $\Gamma(\e)$. The only possibility of getting a nonzero result is linked, as usual, to correlators characterized by IR linearly divergent integrals (without the exponential $e^{-\e t}$). The question we have to ask is whether there are `enough' such states. We would like to show in the sequel that they are plentiful.

Consider states created by multiple products of the factor $H(\phi_u,\e)=\frac 1{K+\phi_u+\e} (\phi_u-\delta \phi_u)$ and contract them with $\Gamma(\e)$. More precisely, let us define
\be
{\mathbf \Psi}_n(\phi_u,\e)= H(\phi_u,\e)^{n-1}   Bc\partial c H(\phi_u,\e)Bc\partial c B,\quad\quad n\geq 2\label{Psin}
\ee
Contracting with $\Gamma(\e)$: $\langle {\mathbf \Psi}_n(\phi_u,\e),\Gamma(\e)\rangle$, we obtain a correlator
whose IR and UV behaviour (before the  $e^{-\e \partial_\e}\e$ operator is applied) is not hard to guess. 
The correlators take the form 
\be
\int_0^\infty ds\, s^n e^{-\tilde \eta s}\,g(s)\int \prod_{i=1}^ndx_i\,\EE, \left( \left(-\frac {\del g(s)}{g(s)}\right)^{n+1} + \ldots+ \left(-\frac {\del g(s)}{g(s)}\right)^{n-k+1} G_s^k +\ldots+ G_s^{n+1}\right)\label{Hn+1}
\ee
where the notation is the same as in section 1.1 ($s=2uT$), but we have tried to make it as compact as possible.
The angular variables $x_i$ have been dropped in $\EE$ and $G_s$ (see, for instance, (\ref{Efinal}) where they are explicitly written down). Using the explicit form of $G_s$ (\ref{Gutheta}), expanding the latter with the binomial formula and integrating over the angular variables, one gets
\be 
\int \prod_{i=1}^ndx_i\,\EE\,G_s^k = \sum_{l=0}^k \frac 1{s^{k-l}} \, \sum_{n_1,\ldots,n_{l}}\,
\frac {P_l(n_1,\ldots,n_{l})}{Q_l(n_1,\ldots,n_{l})} \prod_{i=1}^l \frac 1{p_i(n_1,\ldots,n_{l})+s}\label{intGs}
\ee
the label $l$ counts the number of cosine factors in each term.
Here $n_i$ are positive integral labels which come from the discrete summation in $G_s$; $p_i(n_1,\ldots,n_{l})$ are polynomials linear in $n_i$. Next, $P_l$ and $Q_l$ are polynomials in $n_i$
which come from the integration in the angular variables. Every integration in $x_i$ increases by 1 the difference in the degree of $Q_l$ and $P_l$, so that generically ${\rm deg} Q_l-{\rm deg} P_l= n$. But in some subcases the integration over angular variables give rise to Kronecker deltas among the indices, which may reduce the degree of $Q_l$. So actually the relation valid in all cases is  ${\rm deg} Q_l\geq {\rm deg} P_l$, but one has to take into account that the number of angular variables to be summed over decreases accordingly. 

We are now in the condition to analyze the UV behaviour of (\ref{Hn+1}).
Let us consider, for instance, the first piece  
\be 
\sim \int_0^\infty ds\,e^{-\tilde \eta s} s^{n}\,g(\frac s2) \left( \frac {\del_s g(\frac s2)}{g(\frac s2)}\right)^{n+1}\label{piece}
\ee
Since in the UV $g(\frac s2) \approx \frac 1{\sqrt s}$, it is easy to see that the UV behaviour of the overall integrand is $\sim s^{-\frac 32}$, independently of $n$. As for the other terms, let us consider in the RHS of (\ref{intGs}) the factor that multiplies $\frac 1{s^{k-l}}$ (for $l\geq 2$). Setting $s=0$, the summation
over $n_1,\ldots,n_{l-1}$ is always convergent, so that the UV behaviour of each term in the summation is given by the factor $\frac 1{s^{k-l}}$, with $2\leq l\leq k$. It follows that the most UV divergent term corresponds to $l=0$, $\sim \frac 1{s^k}$. Since in (\ref{Hn+1}) this is multiplied by 
\be
s^n \,g(\frac s2 )\left(-\frac {\del g(\frac s2)}{g(\frac s2)}\right)^{n-k+1}\label{rest}
\ee
we see that the UV behaviour of the generic term in (\ref{Hn+1}) is at most as singular as $\sim s^{-\frac 32}$.
{\it In conclusion the states ${\mathbf \Psi}_n$, when contracted with $\Gamma(\e)$, give rise to the same kind of UV singularity $\sim s^{-\frac 32}$}. Now, for any two such  states, say
${\mathbf \Psi}_n$ and  $ {\mathbf \Psi}_{n'}$, we can form a suitable combination such that the UV singularity cancels. In this way we generate infinite many states, say ${\mathbf \Phi}_n$, which, when contracted
with $\Gamma(\e)$, give rise to UV convergent correlators. 

Let us consider next the IR properties ($s\gg 1$). All the correlators contain the factor $e^{-\tilde \eta s}$ which renders them IR convergent, but we have learnt that the crucial IR properties (in the limit $\e\to 0$) are obtained by ignoring this exponential factor. So, in analyzing the IR properties we will ignore this factor. The first term (\ref{piece}) is very strongly convergent in the IR, because $\del_s g(\frac s2)\approx \frac 1{s^2}$, while $g(\frac s2) \to 1$. For the remaining terms let us consider in the RHS of (\ref{intGs}) the factor that multiplies $\frac 1{s^{k-l}}$ (for $l\geq 2$). To estimate the IR behaviour it is very important to know the degree difference between the polynomials $Q_l$ and $P_l$. Above we said that this difference is always nonnegative. In principle it could vanish, but from the example with   $n=2$, see \cite{BGT1}, we know that there are cancellations and that in fact the difference in degree is at least 2.
If this is so in general, we can conclude that the IR behaviour of the summation in the RHS of (\ref{intGs})
with fixed $l$ is $\sim \frac 1{s^l}$. However, in order to prove such cancellations, one would have to do detailed calculations, which we wish to avoid here. So we will take the pessimistic point of view and assume that, at least for some of the terms, ${\rm deg} Q_l={\rm deg} P_l$ (in which case there remains only one angular integration). In this case the IR behaviour of the corresponding term cannot decrease faster than $\sim \frac 1{s^{l-1}}$. This has to be multiplied by
 $\sim \frac 1{s^{k-l}}$ and by the IR behaviour of (\ref{rest}). This means that the least convergent term
with fixed $k$ in(\ref{intGs}) behaves as $\sim \frac 1{s^{n-k+1}}$. Since $k\leq n+1$, we see that in the worst
hypothesis in the integral (\ref{Hn+1}) there can be linearly divergent terms, before the  $e^{-\e \partial_\e}\e$ operator is applied. If this is so the UV converging ${\mathbf \Phi}_n$ states are not good test states. However we can repeat for the IR singularities what we have done for the UV ones. Taking suitable differences of the ${\mathbf \Phi}_n$'s (this requires a two steps process, first for the linear and then for the logarithmic IR singularities\footnote{In the, so far not met, case where a $\log s$ asymptotic contribution appears in the integrand one would need a three step subtraction process.}), we can create an infinite set of states, ${\mathbf \Omega}_n$, which, when
contracted with $\Gamma(\e)$, yield, before the application of  $e^{-\e \partial_\e}\e$, a finite result. Upon applying $e^{-\e \partial_\e}\e$ they of course vanish. 
These are therefore good (and nontrivial) test states and, on applying $e^{-\e \partial_\e}\e$, they
give 0, i.e. such ${\mathbf \Omega}_n$ annihilate $\Gamma(\e)$.  

We remark that in eq.(\ref{Psin}) the presence of $\e$ in $H(\phi_u,\e)$ is not essential, because in estimating the IR behaviour we have not counted the $e^{-\tilde \eta s}$ factor. Using $\frac 1{K+\phi_u}$ everywhere instead of $\frac 1{K+\phi_u+\e}$, would lead to the same results. This means that
contracting the $\Omega_n$ states among themselves (keeping the same ghost factor) leads to finite correlators {\it with or without $\e$}. This, together with the property of annihilating $\Gamma(\e)$, is a distinctive feature of good test states.

The $\Omega_n(\phi_u,0)$ are however only a first set of good test states. One can envisage a manifold of other such states. Let us briefly describe them, without going into too many details. For instance, let us start
again from (\ref{Psin}) and replace the first $H(\phi_u,0)$ factor with $\frac 1{K+\phi_u+\e} uX^{2k}$ (the term $\delta \phi$ can be dropped). In this way we obtain a new state depending on a new integral label $k$.
However replacing $X^2$ with $X^{2k}$ is a too rough operation which renders the calculations unwieldy, because it breaks the covariance with respect to the rescaling $z \to \frac zt$. It is rather easy to remedy by studying the conformal transformation of $X^{2k}$. The following corrected replacements will do:
\be
uX^2 &\to& u\left(X^2 +2 (\log u+\gamma)\right)=\phi_u\equiv\phi_u^{(1)} \0\\
uX^4 &\to&u\left( X^4 +12 (\log u+\gamma) X^2+ 12 (\log u+\gamma)^2\right)\equiv \phi_u^{(2)}\0\\
&\ldots&\0\\
uX^{2k} &\to&u\left( \sum_{i=0}^k \frac {(2k)!}{(2k-2i)! i!} \left(\log u+\gamma\right)^i X^{2k-2i}\right)\equiv \phi_u^{(k)}\label{X2k}
\ee
The role of the additional pieces on the RHS is to allow us to reconstruct the derivatives of $g(s)$ in computing the correlators, as was done in \cite{BMT}. 

Now let us denote by ${\mathbf \Psi}_n^{(k)}$ the $n$-th state (\ref{Psin}) where $\phi_u-\delta\phi_u$ in the first $H(\phi_u,0)$ factor is replaced by $\phi_u^{(k)}$. Contracting it with $\Gamma(\e)$ it is not hard to see
that the term (\ref{piece}) will be replaced by
\be 
\sim \int_0^\infty ds\,e^{-\tilde \eta s} \, s^{n}\,g(\frac s2) \left( \frac {\del_s g(\frac s2)}{g(\frac s2)}\right)^{n+k}\label{piecenew}
\ee
with analogous generalizations for the other terms. It is evident from (\ref{piecenew}) that the UV behaviour becomes more singular with respect to (\ref{piece}) while the IR one becomes more convergent. This is a general 
property of all the terms in the correlator. Thus fixing $k$ we will have a definite UV singularity, the same up
to a multiplicative factor for all ${\mathbf \Psi}_n^{(k)}$. Therefore by combining a finite number of them we can eliminate the UV singularity and obtain another infinite set of UV convergent states ${\mathbf \Phi}_n^{(k)}$ for any $k$. In general they will be IR convergent (IR subtractions may be necessary for $k=2$).

It goes without saying that the previous construction can be further generalized by replacing in (\ref{Psin}) more than one $X^2$ factors with higher powers $X^{2k}$.

Let us end this section by suggesting another set of states that may be used in order
to construct good test states with a subtraction procedure as above.  Let us consider states containing a certain number of derivatives of $\phi_u$  
\be
{\mathbf \Psi}_{n,k}(\phi_u,\e)= \frac 1{(K+\phi_u+\e)}\frac{1}{(2u)^k}\partial^k\phi_u\, {\mathbf \Psi}_{n-1}(\phi_u,\e)\label{Psink}
\ee
By contracting them with $\Gamma(\e)$ we obtain correlators that, before applying $e^{-\e \partial_\e}\e$,
are defined by integrands in which the UV singularities
are worse (and depend on $k$), while the IR seem to improve by a factor $\sim \frac 1{s^{k-1}}$ with respect
to ${\mathbf \Psi}_n$. However
the derivative $\partial^k$, hitting the propagator $G_s$, increases the degree of $P_l$. The two effects
seem eventually to compensate each other, but the exact IR asymptotic behaviour is more difficult to analyze
in this case, unlike the previous examples. For this reason we leave these states as a suggestion to be analyzed in the future.

\section{Spurious terms: comments and conclusions}

Let us summarize the results we have found.
We think we have abundantly shown in section 7 that the term  $ \Gamma(\e)\equiv e^{-\e \partial_\e} \frac {\e}{K+\phi_u+\e}(\phi_u-\delta \phi_u)c\del c $ in (\ref{RHS1},\ref{B4}), when inserted
in correlators, is either identically vanishing or ambiguous. The first case occurs when it is inserted in a regular correlator, i.e. in a correlator which is convergent even when the factor $e^{-\e t}$ coming from the Schwinger representation of $ \frac 1{K+\phi_u+\e}$ is replaced by 1, which implies that the resulting integral (with the $e^{-\e t}$ factor) is continuous at $\e=0$).  This shows that our calculation of the energy in \cite{BGT1,BGT2} is not affected by the term $ \Gamma(\e)$, as one might have feared (see section 3). That also means that the Schwinger representation of an inverse is correct, provided it is used in the correct way.  

The second case is when the correlator is
at least linearly divergent in the IR (meaning that the correlator is divergent when the factor $e^{-\e t}$ coming from the Schwinger representation of $ \frac 1{K+\phi_u+\e}$ is removed, which implies that the resulting integral (with the $e^{-\e t}$ factor) is discontinuous at $\e=0$) : the typical situation is represented by eq.(\ref{Qenergy3}). In this case we need an {\it ad hoc} prescription in order to extract a finite value from the integral, finite value which is originated, as we have shown, by multiplying a zero by $\infty$. It is clear that this is not the right way to compute the energy of anything (neither solutions, nor non-solutions) 

The formal presence of the term $ \Omega_u^\infty$
in the RHS of (\ref{Schwrong}) or of $e^{-\e \partial_\e} \frac {\e}{K+\phi_u+\e}$ in the RHS of (\ref{IDDu1})  is simply the spy of the fact that we are evaluating the identity (\ref{IDDu}) on a discontinuous correlator. If the correlator's integrand is convergent enough any such addition as $\frac 1{K+\phi_u} \Omega_u^\infty$ is irrelevant and $\frac 1{K+\phi_u}$ is correctly represented by (\ref{Schright}). The appearance of $ \Omega_u^\infty$ or $e^{-\e \partial_\e} \frac {\e}{K+\phi_u+\e}$ becomes a pathology of the Schwinger representation which 
may show up if the problem is not formulated in the proper setting. The appropriate setting is that of distribution theory. In this framework the spurious terms are identically vanishing and there are no violations
of the equation of motion.

All these conclusions are based on explicit evaluations and are unquestionable. This said, it would be nice 
to have a general framework for these problems, a formalization of the rules and procedures we have used above that can be applied in general. At the moment, to our best knowledge, the latter does not exist. The analogy with the case illustrated in section 4 has been instrumental
in understanding the nature of the lump solution problem; the treatment there was based on the theory of distributions. We do not seem to have an analogous theory in the case of string fields, but no doubt this is the right instrument we need in order to treat the singularity problems inherent in the search for solutions in SFT.

We cannot hope to solve this problem here. But we think we have clarified the issue at least on one example
(the relevant example for our present purposes), that is $ \Gamma(\e)$.  An ordinary distribution is just a linear continuous functional on a space of test functions. We can heuristically extend this definition to string fields. A string field distribution is a linear functional on the space of test string fields. In the previous section we have introduced a large set of test states. They are well defined and contain as a particular
case the good test states mentioned before. When $ \Gamma(\e)$ is evaluated on them it gives 0.  Therefore
in distribution theory this expression is identically vanishing. Said otherwise, it is correct to identify $ \Gamma(\e)$ with the zero in distribution theory.

Invoking distribution theory in order to get rid of the spurious terms in the equation of motion (and elsewhere)
may seem {\it ad hoc} at first sight, but the interpretation in terms of distribution theory provides a consistent regularization we need in order to make sense of ambiguities. As we have pointed out in section 4, this is a familiar procedure in theoretical physics in order to carefully define various physical solutions. Apart from the example in section 4, brane solutions in supergravity are often characterized by a metric that explodes when we approach the brane location in the transverse direction, as it depends on some negative power of $r$, $r$ being the transverse distance. However the relevant physical quantities, like the energy density, are finite. There is only one way to give an unambiguous meaning to such solutions: it is to interpret them in the framework of distribution theory. 

A formalization of the idea of string field distribution (beyond the example of $\Gamma(\e)$ studied in detail above) is possible, but, as we pointed out above, to our best knowledge the relevant formalism has not been developed so far. Perhaps the right mathematical setting is offered by the vector distribution theory. The theory of vector distributions was developed by Laurent Schwartz,  \cite{Schwartz}. The basic objects are a topological vector space and the space of test functions. A distributions is a linear continuous map from the latter to the former. More practically we can think of test vector functions as tensor products of ordinary scalar test functions by vectors and a vector distribution as a space dependent vector, while the evaluation on a vector test function is the ordinary scalar product followed by an ordinary integration. In our case the expression $\frac 1{K+\phi_u}$ should be regarded as a vector distribution. It goes without saying that much work has to be done in order to clarify definitions and show applicability of such formalism in the context of SFT.

{\bf Acknowledgments} 

L.B would like to thank Nobuyuki Ishibashi and Branko Dragovic for useful discussions.
The work of L.B. and S.G. was supported in part by the MIUR-PRIN contract 2009-KHZKRX. 
D.D.T. would like to thank SISSA for the kind hospitality during part of this research. The work of D.D.T. was supported by the Korean Research Foundation Grant funded by the Korean Government
with grant number KRF 2009-0077423.

While this new version of the paper was in preparation three papers \cite{MS,EM2,EM3} appeared whose content has links with the subjects discussed here.

\vspace{0.5cm}

\section*{Appendix}
\appendix
 
\section{$\Delta_\ve^{(1)}$ and  $\Delta_\ve^{(2)}$ }

This appendix is devoted to the analytical elaboration of the terms 
$\Delta_\ve^{(1)}$ and $\Delta_\ve^{(2)}$ with the aim to find the most suitable form for their numerical evaluation.

\subsection{$\Delta_\ve^{(1)}$}

$\Delta_\ve^{(1)}$ is the difference between the two terms
\be
\langle\psi_u,\psi_u\psi_u\rangle&=&- \int_0^\infty ds\;
s^2\int_0^1 dy\int_0^{y} dx\, {\cal
E}(1-y,x) \,\g(s)\label{Fsu}\\
&&\cdot\Bigg\{ \Big(-\frac{\partial_{s}\g(s)}{\g(s)}\Big)^3
 +G_{s}(2\pi x)G_{s}(2\pi(x-y))G_{s}(2\pi y)\0\\
&&\,-\frac 12\frac{\partial_{s}\g(s)}{\g(s)}\Big(G_{s}^2(2\pi x)+G_{s}^2(2\pi
y)+G_{s}^2(2\pi (x-y))\Big) \Bigg\}.\0
\ee
and
\be
\langle\psi_u^\ve,\psi_u^\ve \psi_u^\ve\rangle&=&- \int_0^\infty ds\;
s^2\int_0^1 dy\int_0^{y} dx\,e^{-\eta s}{\cal
E}(1-y,x) \, \g(s)\label{Fs3epsi}\\
&&\cdot\Bigg\{\Big(\eta -\frac{\partial_{s}\g(s)}{\g(s)}\Big)^3 
 +G_{s}(2\pi x)G_{s}(2\pi(x-y))G_{s}(2\pi y)\0\\
&&\,+\frac 12\Big(\eta-\frac{\partial_{s}\g(s)}{\g(s)}\Big)\Big(G_{s}^2(2\pi(x-y))+  G_{s}^2(2\pi x)+G_{s}^2(2\pi y)\Big) \Bigg\}.\0
\ee
We will rewrite $\Delta_\ve^{(1)}$  as follows:
\be
\Delta_\ve^{(1)}&=& \langle\psi_u,\psi_u\psi_u\rangle-\langle\psi_u^\ve,\psi_u^\ve \psi_u^\ve\rangle\label{Delta1}\\
&=& \int_0^\infty ds\;s^2\int_0^1 dy\int_0^{y} dx\,e^{-\eta s}{\cal E}(1-y,x)\left\{ \left(1-e^{-\eta s}\right)\left[
\Big(-\frac{\partial_{s}\g(s)}{\g(s)}\Big)^3 \right.\right.\0\\
&&-\frac 12\Big(\frac{\partial_{s}\g(s)}{\g(s)}\Big)\Big(G_{s}^2(2\pi(x-y))+  G_{s}^2(2\pi x)+G_{s}^2(2\pi y)\Big)\0\\
&&+ G_{s}(2\pi x)G_{s}(2\pi(x-y))G_{s}(2\pi y)\bigg{]}\0\\
&&- e^{-\eta s} \left[ \eta^3-3\eta^2 \frac{\partial_{s}\g(s)}{\g(s)}+3\eta  \Big(\frac{\partial_{s}\g(s)}{\g(s)}\Big)^2\right. \0\\
&&+\left.\left. \frac 12 \eta \Big(G_{s}^2(2\pi(x-y))+  G_{s}^2(2\pi x)+G_{s}^2(2\pi y)\Big)\right]\right\}
\0 
\ee
The angular integrations are the same as in \cite{BGT1}. Using these results one arrives at the
formulas in {\bf definitions.nb} and can proceed to their numerical evaluation. In many cases this means a numerical integration over $s$ from 0 to $\infty$. This can be done straightaway or by splitting the integration from 0 to 1 and from 1 to $\infty$. In some cases it is not possible to numerically integrate up to $\infty$. Then one proceeds to integrate up to a finite number, say one million, and verify that it is large enough so that any increase will lead to irrelevant contributions for the degree of accuracy we want. In the same way one proceeds with the discrete summations when their analytic resummation is impossible. The numerical results 
can be found in Table 1.

\subsection{$\Delta_\ve^{(2)}$}

$\Delta_\ve^{(2)}$ is the difference between the two terms
\be
\langle\psi_u^\ve,\psi_u\psi_u\rangle&=&- \int_0^\infty ds\;
s^2\int_0^1 dy\int_0^{y} dx\,e^{-\eta s}{\cal
E}(1-y,x) \,e^{\eta sy}\, \g(s)\label{Fs2}\\
&&\cdot\Bigg\{\Big(\eta -\frac{\partial_{s}\g(s)}{\g(s)}\Big)\Big(-\frac{\partial_{s}\g(s)}{\g(s)}\Big)^2
 +G_{s}(2\pi x)G_{s}(2\pi(x-y))G_{s}(2\pi y)\0\\
&&\,+\frac 12\Big(\eta-\frac{\partial_{s}\g(s)}{\g(s)}\Big)G_{s}^2(2\pi x)+ \frac 12\Big(-\frac{\partial_{s}\g(s)}{\g(s)}\Big)\Big(G_{s}^2(2\pi
y)+G_{s}^2(2\pi (x-y))\Big) \Bigg\}.\0
\ee
and
\be
\langle\psi_u,\psi_u^\ve\psi_u^\ve\rangle&=&- \int_0^\infty ds\;
s^2\int_0^1 dy\int_0^{y} dx\,e^{-\eta s}{\cal
E}(1-y,x) \,e^{\eta sx}\, \g(s)\label{Fs3}\\
&&\cdot\Bigg\{\Big(\eta -\frac{\partial_{s}\g(s)}{\g(s)}\Big)^2\Big(-\frac{\partial_{s}\g(s)}{\g(s)}\Big)
 +G_{s}(2\pi x)G_{s}(2\pi(x-y))G_{s}(2\pi y)\0\\
&&\,+\frac 12\Big(-\frac{\partial_{s}\g(s)}{\g(s)}\Big)G_{s}^2(2\pi(x-y))+ \frac 12\Big(\eta-\frac{\partial_{s}\g(s)}{\g(s)}\Big)\Big(G_{s}^2(2\pi
x)+G_{s}^2(2\pi y)\Big) \Bigg\}.\0
\ee
where $\eta= \frac {\ve}{2 s}$.

The first step consists in rewriting these terms
as far as possible as integrals of both $x$ and $y$ between 0 and 1 in order to simplify the angular integrations. In fact any integral $\int_0^1 dy \int_0^y dx\, f(x,y)$ can be rewritten as  $\frac 12 \int_0^1 dy \int_0^1 dx\, f(x,y)$ provided $f(x,y)=f(y,x)$.

Using this fact, after some work one gets
\be 
&&\Delta^{(2)}_\ve \equiv\langle\psi_u^\ve,\psi_u\psi_u\rangle- \langle\psi_u,\psi_u^\ve\psi_u^\ve\rangle=\label{epsi1}\\
&=&-\frac 12\int_0^\infty ds\,s^2\int_0^1dy \int_0^1 dx\left( e^{\eta s y}-e^{\eta x}\right){\cal E}(1-y,x)\, H(x,y,\eta,s) \0\\
&&- \frac {\eta}2 \int_0^\infty ds\,s^2\int_0^1dy \int_0^1 dx\left( e^{\eta s y}-e^{\eta x}\right){\cal E}(1-y,x)\,\g(s)\, e^{-\eta s} \left(- \frac {\partial_s \g(s)}{\g(s)}\right)^2\0\\
&&- \frac {\eta}4\int_0^\infty ds\,s^2\int_0^1dy \int_0^1 dx\left( e^{\eta s y}G^2_s(2\pi x)-e^{\eta s x}G_s^2(2\pi y)\right){\cal E}(1-y,x)\,\g(s)\, e^{-\eta s}\0\\
&&-\int_0^\infty ds\,s^2\int_0^1dy \int_0^y dx {\cal E}(1-y,x)\,\g(s)\, e^{-\eta s}\, 
e^{\eta s x}\0\\
&&\quad\quad \cdot\left( \eta^2 \left( \frac {\partial_s \g(s)}{\g(s)}\right)
- {\eta} \left(\frac {\partial_s \g(s)}{\g(s)}\right)^2-\frac {\eta}2 G^2_s(2\pi x)\right)\0
\ee
where
\be 
H(x,y,\eta,s)&=&\g(s) e^{-\eta s}\Bigg\{\Big( -\frac{\partial_{s}\g(s)}{\g(s)}\Big)^3 
 +G_{s}(2\pi x)G_{s}(2\pi(x-y))G_{s}(2\pi y)\label{epsi2}\\
&&\,+\frac 12\Big(-\frac{\partial_{s}\g(s)}{\g(s)}\Big)\Big(G_{s}^2(2\pi(x-y))+  G_{s}^2(2\pi x)+G_{s}^2(2\pi y)\Big) \Bigg\} \0
\ee
Simplifying further
 \be 
&&\Delta^{(2)}_\ve=\langle\psi_u^\ve,\psi_u\psi_u\rangle- \langle\psi_u,\psi_u^\ve\psi_u^\ve\rangle=\label{epsi3}\\
&=&-\int_0^\infty ds\,s^2\int_0^1dy \int_0^1 dx\, e^{\eta s y}\,{\cal E}(1-y,x)\, H(x,y,\eta,s) \0\\
&&-  {\eta} \int_0^\infty ds\,s^2\int_0^1dy \int_0^1 dx\, e^{\eta s y}\,{\cal E}(1-y,x)\,\g(s)\, e^{-\eta s} \left(- \frac {\partial_s \g(s)}{\g(s)}\right)^2\0\\
&&- \frac {\eta}2\int_0^\infty ds\,s^2\int_0^1dy \int_0^1 dx \, e^{\eta s y}\,G^2_s(2\pi x) {\cal E}(1-y,x)\,\g(s)\, e^{-\eta s}\0\\
&&-\int_0^\infty ds\,s^2\int_0^1dy \int_0^y dx\, {\cal E}(1-y,x)\,\g(s)\, e^{-\eta s}\,  
e^{\eta s x}\0\\
&&\quad\quad \cdot\left( \eta^2 \left( \frac {\partial_s \g(s)}{\g(s)}\right)
- {\eta} \left(\frac {\partial_s \g(s)}{\g(s)}\right)^2-\frac {\eta}2 G^2_s(2\pi x)\right)\0
\ee
Notice that all the terms in (\ref{epsi3}) are UV finite and there is no need for UV subtractions.

Next we integrate over $x$ and $y$. The results are summarized in the sequel.

\subsubsection{Terms of $\Delta_\ve^{(2)}$ ont containing  $G_s$}

The terms not containing $G_s$ sum up to
\be 
&-&\int_0^\infty ds\,s^2\int_0^1dy \int_0^1 dx\, e^{\eta s y}\,{\cal E}(1-y,x) \,g(s) \,e^{-\eta s}
\left[\left(-\frac{\partial_{s}\g(s)}{\g(s)}\right)^3 +\eta \left(-\frac{\partial_{s}\g(s)}{\g(s)}\right)^2\right]\0\\
&-&\int_0^\infty ds\,s^2\int_0^1dy \int_0^y dx\, e^{\eta s x}\,{\cal E}(1-y,x) \,g(s) \,e^{-\eta s}
\left[\eta^2\left(\frac{\partial_{s}\g(s)}{\g(s)}\right) -\eta \left(-\frac{\partial_{s}\g(s)}{\g(s)}\right)^2\right]\0\\
&=&  \int_0^\infty ds\, {\cal O}(s,\eta)\label{Oseta}
\ee
where
\be
&&{\cal O}(s,\eta)=s^2 \frac{2-2 e^{-s \eta }}{4 \pi ^2+s^2 \eta ^2} \left[\left(-\frac{\partial_{s}\g(s)}{\g(s)}\right)^3 +\eta \left(-\frac{\partial_{s}\g(s)}{\g(s)}\right)^2 \right]\0\\
&&+s^2\frac{2 e^{-s \eta } \left(-8 \pi ^2+8 e^{s \eta } \pi ^2+4 \pi ^2 s \eta +s^3 \eta ^3\right)}{s \eta  \left(4 \pi ^2+s^2 \eta ^2\right)^2}\left[\eta^2\left(\frac{\partial_{s}\g(s)}{\g(s)}\right) -\eta \left(-\frac{\partial_{s}\g(s)}{\g(s)}\right)^2\right]\label{Oseta1}
\ee

\subsubsection{Terms of $\Delta_\ve^{(2)}$ containing $G_s^2$}

In terms containing $G_s^2$ we come across the following angular integrals
\be
&& \quad\quad\int_0^1dy \int_0^1 dx\, e^{\eta s y}\,{\cal E}(1-y,x) \, e^{-\eta s}=- 2 \frac {1-e^{-s\eta}}{4\pi^2+s^2\eta^2}\label{Gssquare}
\0\\
&& \quad\quad\int_0^1dy \int_0^1 dx\, e^{\eta s y}\,{\cal E}(1-y,x) \, e^{-\eta s}\cos 2\pi k x=\delta_{k,1}\frac {1-e^{-s\eta}}{4\pi^2+s^2\eta^2}\0\\
&& \quad\quad\int_0^1dy \int_0^1 dx\, e^{\eta s y}\,{\cal E}(1-y,x) \, e^{-\eta s}\cos 2\pi k y= c(s,\eta,k)\0\\
&& \quad\quad\int_0^1dy \int_0^1 dx\, e^{\eta s y}\,{\cal E}(1-y,x) \, e^{-\eta s}\cos 2\pi k(x-y)= \delta_{k,1}\frac {1-e^{-s\eta}}{4\pi^2+s^2\eta^2} \0\\
&& \quad\quad\int_0^1dy \int_0^1 dx\, e^{\eta s y}\,{\cal E}(1-y,x) \, e^{-\eta s}\cos 2\pi k x \cos 2\pi n x = \0\\
&& \quad\quad =  \frac {1-e^{-s\eta}}{4\pi^2+s^2\eta^2} \left( -\delta_{k,n}+\frac 12 \delta_{k,n+1}+ \frac 12
\delta_{k,n-1}\right)\0\\
&& \quad\quad\int_0^1dy \int_0^1 dx\, e^{\eta s y}\,{\cal E}(1-y,x) \, e^{-\eta s}\cos 2\pi k y \cos 2\pi n y =
f(s,\eta,k,n) \0\\
&& \quad\quad\int_0^1dy \int_0^1 dx\, e^{\eta s y}\,{\cal E}(1-y,x) \, e^{-\eta s}\cos 2\pi k (x-y) \cos 2\pi n (x-y) = \0\\
&&\quad\quad =  \frac {1-e^{-s\eta}}{4\pi^2+s^2\eta^2} \left( -\delta_{k,n}+\frac 12 \delta_{k,n+1}+ \frac 12
\delta_{k,n-1}\right)\0
\ee
The coefficients $c(s,\eta,k)$ and $f(s,\eta,k,n)$ can be found in the Mathematica file {\bf definitions.nb}.

After these angular integrations the terms quadratic in $G_s$ become
\be 
&&- \frac {\eta}2\int_0^\infty ds\,s^2\int_0^1dy \int_0^1 dx \, e^{\eta s y}\,G^2_s(2\pi x) {\cal E}(1-y,x)\,\g(s)\, e^{-\eta s}\0\\
&&= {\eta}\int_0^\infty ds\,s^2 \g(s)\,  \frac {1-e^{-s\eta}}{4\pi^2+s^2\eta^2}\left(\frac 1{s^2} -\frac 2{s(s+1)} -\frac 2{s+1}+2\psi^{(1)}(1+s)\right)\label{epsi4}
\ee
(the integrand of (\ref{epsi4}) is denoted $Q1(s,\eta)$ in the Mathematica file)
and
\be
&&-\int_0^\infty ds\,s^2\int_0^1dy \int_0^1 dx\, e^{\eta s y}\,{\cal E}(1-y,x)\,\g(s) \,e^{-\eta s} \,\0\\
&&\quad\quad \cdot\frac 12\Big(-\frac{\partial_{s}\g(s)}{\g(s)}\Big)\Big(G_{s}^2(2\pi(x-y))+  G_{s}^2(2\pi x)+G_{s}^2(2\pi y)\Big) =\0\\
&&= \int_0^\infty ds\,s^2\, \partial_{s}\g(s)\0\\
&&\cdot \left[\frac {1-e^{-s\eta}}{4\pi^2+s^2\eta^2}\left(-\frac 3{s^2} +\frac 4{s(s+1)} +\frac 4{s+1}-4\psi^{(1)}(1+s)\right)\right.\quad\quad\longleftrightarrow Q2a(s,\eta)\0\\
&& \quad+\frac 2s \sum_{k=1}^\infty \frac {c(s,\eta,k)}{k+s} \quad\quad \quad\quad\longleftrightarrow Q2b(s,\eta,k)\0\\ 
&&\quad+ \left.2 \sum_{k,n=1}^\infty \frac {f(s,\eta,k,n)}{(k+s)(n+s)} \right]\quad\quad\longleftrightarrow Q2c(s,\eta,k,n)\label{epsi5}
\ee
In this unorthodox notation $Q2a(s,\eta), Q2b(s,\eta,k), Q2c(s,\eta,k)$ represent the corresponding integrand and summands.

We define also
\be
Q2b(s,\eta)&=&\sum_{k=1}^\infty Q2b(s,\eta,k)\label{Q2b}\\
Q2c(s,\eta,n)&=&\sum_{k=1}^\infty Q2c(s,\eta,k,n)\0
\ee
These summations can be carried out analytically.
 
Finally 
\be 
&& \frac {\eta}2\int_0^\infty ds\,s^2\int_0^1dy \int_0^y dx \, e^{\eta s x}\,G^2_s(2\pi x) {\cal E}(1-y,x)\,\g(s)\, e^{-\eta s}\0\\
&=& \frac{\eta}2 \int_0^\infty ds\,s^2 \g(s)\left[-2 e^{-\eta s} \frac {8\pi^2 (e^{\eta s}-1)+ 4\pi^2 \eta s +s^3\eta^3}
{\eta s^3(4\pi^2+s^2\eta^2)^2}\right. \quad\quad \longleftrightarrow Q3a(s,\eta)\0\\
&&\quad +\frac 4s \sum_{n=1}^\infty a(s,\eta,n) \frac 1{n+s} \quad\quad\quad \quad\quad\quad\longleftrightarrow Q3b(s,\eta,n)\0\\ 
&&\quad \left. +4\sum_{k,n=1}^\infty b(s,\eta,k,n) \frac 1{(k+s)(n+s)} \right] \quad\quad  \quad\quad\quad \longleftrightarrow Q3c(s,\eta,k,n)\label{epsi6}
\ee
The $Q$ symbols refer to the Mathematica file {\bf definitions.nb}, where the definitions of
$a(s,\eta,n)$ and $b(s,\eta,k,n)$ can be found as well.

There are problems with $Q3b(s,\eta)$ for $n=1$. So we split 
\be 
Q3b(s,\eta) \rightarrow Q3b1(s,\eta)+Q3b(s,\eta)\0
\ee
where now 
\be
&& Q3b1(s,\eta)= 2 \eta s \g(s)\frac{ a(s,\eta,1)}{1+s}\0\\
&& Q3b(s,\eta,n)=  2 \eta s \g(s)  a(s,\eta,n) \frac 1{n+s},\quad\quad n\geq 2\0
\ee

We define also
\be
Q3c(s,\eta,n)=\sum_{k=1}^\infty Q3c(s,\eta,k,n)\label{Q3c}
\ee

The relevant definitions are contained in the Mathematica file {\bf definitions.nb}.

\subsubsection{Terms of $\Delta_\ve^{(2)}$ containing $G_s^3$ }

\be
&&-\int_0^\infty ds\,s^2\int_0^1dy \int_0^1 dx\, e^{\eta s y}\,{\cal E}(1-y,x)\, 
\g(s) e^{-\eta s} \,G_{s}(2\pi x)G_{s}(2\pi(x-y))G_{s}(2\pi y)\0\\
&& =-\int_0^\infty ds\,s^2\, \g(s) \left[   \frac {1-e^{-s\eta}}{4\pi^2+s^2\eta^2}\left(-
\frac 2{s^3} +\frac 4{s^2(s+1)}\right) + \frac 2{s^2}\sum_{n=1}^\infty \frac {t(s,\eta,n)}{n+s}\right.\label{epsi7}\\
&&  +\frac 4s \left(\sum_{n=1}^\infty \frac {s0(s,\eta,n)}{(n+s)^2}+ \sum_{n=1}^\infty
\frac {sp(s,\eta,n)}{(n+s)(n+1+s)}+ \sum_{n=2}^\infty \frac {sm(s,\eta,n)}{(n+s)(n-1+s)}\right.\0\\
&&\left. + \frac 1{s+1}\sum_{n=1}^\infty \frac{r1(s,\eta,n)}{n+s} + \frac 1{s+1} \sum_{n=1}^\infty \frac {v(s,\eta,n)}{n+s}\right)\0\\
&&+ 8\sum_{n,k=1}^\infty \frac {w0(s,\eta,k,n)}{(n+s)(k+s)^2} +8 \sum_{n,k=1}^\infty 
\frac {wp(s,\eta,k,n)}{(n+s)(k+s)(k+1+s)}\0\\
&&\left. + 8 \sum_{n=1,k=2}^\infty
\frac{wm(s,\eta,k,n)}{(n+s)(k+s)(k-1+s)}\right] \0
\ee
We have also the relation:
\be 
t(s,\eta,n)=-2 r1(s,\eta,n)=-2 v(s,\eta,n)\0
\ee

All the coefficients $s0(n), sp(n),sm(n), r1(n), r(k,n),v(n),w0(k,n),wp(k,n), wm(k,n)$ are defined in  {\bf definitions.nb}. Let us rearrange the terms as follows
\be
&&-\int_0^\infty ds\,s^2\int_0^1dy \int_0^1 dx\, e^{\eta s y}\,{\cal E}(1-y,x)\, 
\g(s) e^{-\eta s} \,G_{s}(2\pi x)G_{s}(2\pi(x-y))G_{s}(2\pi y)\0\\
&& =-\int_0^\infty ds\,s^2\, \g(s) \left[   \frac {1-e^{-s\eta}}{4\pi^2+s^2\eta^2}\left(-
\frac 2{s^3} +\frac 4{s^2(s+1)}\right)\right. \quad\quad\quad\longleftrightarrow C1(s,\eta) \0\\
&& + \frac 2{s^2}\sum_{n=1}^\infty \frac {t(s,\eta,n)}{n+s}   +\frac 4s \left( \frac 1{s+1}\sum_{n=1}^\infty \frac{r1(s,\eta,n)}{n+s}+ \frac 1{s+1} \sum_{n=1}^\infty \frac {v(s,\eta,n)}{n+s}\right)\quad \longleftrightarrow C2(s,\eta) \0\\
&&  +\frac 4s \left(\sum_{n=1}^\infty \frac {s0(s,\eta,n)}{(n+s)^2}+ \sum_{n=1}^\infty
\frac {sp(s,\eta,n)}{(n+s)(n+1+s)}+ \sum_{n=2}^\infty \frac {sm(s,\eta,n)}{(n+s)(n-1+s)}\right) \longleftrightarrow C3(s,\eta,n) \0\\
&&+ 8\sum_{n,k=1}^\infty \frac {w0(s,\eta,k,n)}{(n+s)(k+s)^2}\quad\quad\quad\quad  \longleftrightarrow C41(s,\eta,k,n)  \0\\
&&+ 8 \sum_{n,k=1}^\infty 
\frac {wp(s,\eta,k,n)}{(n+s)(k+s)(k+1+s)}\quad\quad \longleftrightarrow C42(s,\eta,k,n) \0\\
&&+ \left. 8 \sum_{n=1,k=2}^\infty
\frac{wm(s,\eta,k,n)}{(n+s)(k+s)(k-1+s)}\right] \quad\quad\longleftrightarrow C43(s,\eta,k,n) \label{cubicCi}
\ee

We may also define
\be
C4i(s,\eta,k)= \sum_{n=1}^\infty C4i(s,\eta,k,n),\quad\quad i=1,2,3 \label{C4iseta}
\ee
The summation can be carried out analytically. But a more effective way from a numerical point of view is to first define
\be
w1(s,\eta,k,n)= wp(s,\eta,k,n)+wm(s,\eta,k+1,n)\0
\ee
and then 
\be
C4a(s,\eta,H,K,M,N)=-8s{}^2 \g(s)\sum_{k=H}^K\sum_{n=M}^N\left(\frac{ w0(s,\eta,k,n)}{(n+s)(k+s)^2}+
 +\frac{ w1(s,\eta ,k,n)}{(n+s) (k+s)(k+1+s)}\right)\0
\ee
To obtain good numerics one has to sum the three central diagonals separately from the rest.

The rest of the calculations (sums and integrals) from now on are carried out numerically.
In Table 3 are some samples of the results for the various types of terms
\begin{table}[ht]
$\begin{matrix}\quad &\eta: & 10\,&1\,&0.5 \, &  0.1\,& 0.05\,\\
\quad&{\rm 0-th\,\, order}:\quad& -0.17701\quad& -0.02851\quad&-0.01504\quad&  -0.00317\quad& -0.00159\\
\quad&{\rm quadratic} :\quad& -0.01578\ \quad& -0.04205\quad& -0.04103\quad&-0.03865\quad&  -0.03886\\
\quad&{\rm cubic} :\quad& 0.17848\ \quad& 0.03970\quad& 0.02274\quad& 0.00614\quad& 0.00433
\end{matrix}$
\caption{Samples of contributions of different terms to $ \Delta^{(2)}_\ve$}
\end{table}


\begin{thebibliography}{99}

\bibitem{BMT}
  L.~Bonora, C.~Maccaferri and D.~D.~Tolla,
 {\it Relevant Deformations in Open String Field Theory: a Simple Solution for
  Lumps,}
 JHEP 1111:107,2011; arXiv:1009.4158 [hep-th].

\bibitem{BGT1}
  L.~Bonora, S.~Giaccari and D.~D.~Tolla,
{\it The energy of the analytic lump solution in SFT,} 
JHEP 08(2011)158. ArXiv:1105.5926 [hep-th]. Erratum: JHEP 04(2012)001 

\bibitem{EM}
  T.~Erler and C.~Maccaferri,
 {\it Comments on Lumps from RG flows,}
  arXiv:1105.6057 [hep-th].

\bibitem{BGT2}
  L.~Bonora, S.~Giaccari and D.~D.~Tolla,
{\it Analytic solutions for Dp branes in SFT,}
JHEP 12(2011)033; arXiv:1106.3914 [hep-th]. 

\bibitem{Witten:1985cc}
  E.~Witten,
  {\it Noncommutative Geometry And String Field Theory,}
  Nucl.\ Phys.\  B {\bf 268} (1986) 253.

\bibitem{Sen:1999mh}
  A.~Sen,
  ``Descent relations among bosonic D-branes,''
  Int.\ J.\ Mod.\ Phys.\  A {\bf 14}, 4061 (1999)
  [arXiv:hep-th/9902105].

\bibitem{Sen:1999xm}
  A.~Sen,
  ``Universality of the tachyon potential,''
  JHEP {\bf 9912}, 027 (1999)
  [arXiv:hep-th/9911116].

\bibitem{Schnabl05}
  M.~Schnabl,
  {\it Analytic solution for tachyon condensation in open string field theory,}
  Adv.\ Theor.\ Math.\ Phys.\  {\bf 10} (2006) 433
  [arXiv:hep-th/0511286].

\bibitem{Okawa1}
  Y.~Okawa,
 {\it Comments on Schnabl's analytic solution for tachyon condensation in
  Witten's open string field theory,}
  JHEP {\bf 0604} (2006) 055
  [arXiv:hep-th/0603159].


\bibitem{ErlerSchnabl}
  T.~Erler and M.~Schnabl,
  {\it A Simple Analytic Solution for Tachyon Condensation,}
  arXiv:0906.0979 [hep-th].

\bibitem{RZ06}
  L.~Rastelli and B.~Zwiebach,
  {\it Solving open string field theory with special projectors,}
arXiv:hep-th/0606131.

\bibitem{ORZ}
  Y.~Okawa, L.~Rastelli and B.~Zwiebach,
 {\it Analytic solutions for tachyon condensation with general projectors,}
  arXiv:hep-th/0611110.

\bibitem{Fuchs0}
  E.~Fuchs and M.~Kroyter,
 {\it On the validity of the solution of string field theory,}
  JHEP {\bf 0605} (2006) 006
  [arXiv:hep-th/0603195].

\bibitem{Erler:2006hw}
  T.~Erler,
  {\it Split string formalism and the closed string vacuum},
  JHEP {\bf 0705}, 083 (2007)
  [arXiv:hep-th/0611200].

\bibitem{Erler:2006ww}
  T.~Erler,
  {\it Split string formalism and the closed string vacuum. II},
  JHEP {\bf 0705}, 084 (2007)
  [arXiv:hep-th/0612050].

\bibitem{Erler:2007xt}
  T.~Erler,
  {\it Tachyon Vacuum in Cubic Superstring Field Theory,}
  JHEP {\bf 0801} (2008) 013
  [arXiv:0707.4591 [hep-th]].


\bibitem{Arroyo:2010fq}
  E.~A.~Arroyo,
  {\it Generating Erler-Schnabl-type Solution for Tachyon Vacuum in Cubic
  Superstring Field Theory},
  arXiv:1004.3030 [hep-th].


\bibitem{Zeze:2010jv}
  S.~Zeze,
  {\it Tachyon potential in KBc subalgebra},
  arXiv:1004.4351 [hep-th].

\bibitem{Zeze:2010sr}
  S.~Zeze,
  {\it Regularization of identity based solution in string field theory},
  arXiv:1008.1104 [hep-th].

\bibitem{Arroyo:2010sy}
  E.~A.~Arroyo,
  {\it Comments on regularization of identity based solutions in string field
  theory,}
  arXiv:1009.0198 [hep-th].

\bibitem{Murata}
  M.~Murata and M.~Schnabl,
  {\it On Multibrane Solutions in Open String Field Theory,}
  Prog.\ Theor.\ Phys.\ Suppl.\  {\bf 188}, 50 (2011)
  [arXiv:1103.1382 [hep-th]].

\bibitem{Ghoshal} 
  D.~Ghoshal,
{\it Fisher Equation for a Decaying Brane,}
  JHEP {\bf 1112}, 015 (2011)
  [arXiv:1108.0094 [hep-th]].

  \bibitem{KORZ}
  M.~Kiermaier, Y.~Okawa, L.~Rastelli and B.~Zwiebach,
 {\it Analytic solutions for marginal deformations in open string field theory,}
  arXiv:hep-th/0701249.

  \bibitem{Schnabl:2007az}
  M.~Schnabl,
 {\it Comments on marginal deformations in open string field theory,}
  arXiv:hep-th/0701248.

\bibitem{Kluson:2003xu}
  J.~Kluson,
{\it Exact solutions in SFT and marginal deformation in BCFT,}
  JHEP {\bf 0312}, 050 (2003).
  [hep-th/0303199].

\bibitem{Kiermaier:2007vu}
  M.~Kiermaier and Y.~Okawa,
  {\it Exact marginality in open string field theory: a general framework,}
  arXiv:0707.4472 [hep-th].


\bibitem{Fuchs3}
  E.~Fuchs, M.~Kroyter and R.~Potting,
  {\it Marginal deformations in string field theory,}
   arXiv:0704.2222 [hep-th].


\bibitem{Lee:2007ns}
  B.~H.~Lee, C.~Park and D.~D.~Tolla,
  {\it  Marginal Deformations as Lower Dimensional D-brane Solutions in Open String
  Field theory,}
  arXiv:0710.1342 [hep-th].


\bibitem{Kwon:2008ap}
  O.~K.~Kwon,
  {\it Marginally Deformed Rolling Tachyon around the Tachyon Vacuum in Open
  String Field Theory},
  Nucl.\ Phys.\  B {\bf 804}, 1 (2008)
  [arXiv:0801.0573 [hep-th]].

  \bibitem{Okawa2}
  Y.~Okawa,
{\it Analytic solutions for marginal deformations in open
superstring field
  theory,}
  arXiv:0704.0936 [hep-th].

\bibitem{Okawa3}
  Y.~Okawa,
{\it Real analytic solutions for marginal deformations in open
superstring field theory,}
  arXiv:0704.3612 [hep-th].





\bibitem{Kiermaier:2007ki}
  M.~Kiermaier and Y.~Okawa,
  {\it General marginal deformations in open superstring field theory,}
  arXiv:0708.3394 [hep-th].



\bibitem{Erler:2007rh}
  T.~Erler,
  {\it Marginal Solutions for the Superstring,}
  JHEP {\bf 0707} (2007) 050
  [arXiv:0704.0930 [hep-th]].


  \bibitem{Fuchs4}
E.~Fuchs and M.~Kroyter, {\it Analytical Solutions of Open String
Field Theory,}
  arXiv:0807.4722 [hep-th].

\bibitem{Schnabl:2010tb}
  M.~Schnabl,
  {\it Algebraic solutions in Open String Field Theory - a lightning review},
  arXiv:1004.4858 [hep-th].
  \bibitem{lumps}
  N.~Moeller, A.~Sen and B.~Zwiebach,
  {\it D-branes as tachyon lumps in string field theory,}
  JHEP {\bf 0008} (2000) 039
  [arXiv:hep-th/0005036].

\bibitem{Witten}
  E.~Witten,
 {\it Some computations in background independent off-shell string theory,}
  Phys.\ Rev.\  D {\bf 47}, 3405 (1993)
  [arXiv:hep-th/9210065].

\bibitem{Kutasov}
  D.~Kutasov, M.~Marino and G.~W.~Moore,
{\it Some exact results on tachyon condensation in string field
theory,}
  JHEP {\bf 0010}, 045 (2000)
  [arXiv:hep-th/0009148].

\bibitem{Ellwood} I.~Ellwood,
  {\it Singular gauge transformations in string field theory,}
  JHEP {\bf 0905}, 037 (2009)
  [arXiv:0903.0390 [hep-th]].

\bibitem{EllwoodSchnabl}
  I.~Ellwood and M.~Schnabl,
 {\it Proof of vanishing cohomology at the tachyon vacuum,}
  JHEP {\bf 0702} (2007) 096
  [arXiv:hep-th/0606142].

\bibitem{RSZ}
  L.~Rastelli, A.~Sen and B.~Zwiebach,
 {\it Star algebra spectroscopy,}
  JHEP {\bf 0203}, 029 (2002)
  [arXiv:hep-th/0111281].

\bibitem{Hata} 
H.~Hata and T.~Kojita,
 {\it Winding Number in String Field Theory,}
  JHEP {\bf 1201}, 088 (2012)
  [arXiv:1111.2389 [hep-th]].

\bibitem{BMST1}
  L.~Bonora, C.~Maccaferri, R.~J.~Scherer Santos and D.~D.~Tolla,
  {\it Ghost story. I. Wedge states in the oscillator formalism,}
  JHEP {\bf 0709}, 061 (2007)
  [arXiv:0706.1025 [hep-th]].

\bibitem{BMST2}
  L.~Bonora, C.~Maccaferri, R.~J.~Scherer Santos and D.~D.~Tolla,
 {\it Ghost story. II. The midpoint ghost vertex,}
  JHEP {\bf 0911}, 075 (2009)
  [arXiv:0908.0055 [hep-th]].

\bibitem{BMT3}
  L.~Bonora, C.~Maccaferri and D.~D.~Tolla,
{\it Ghost story. III. Back to ghost number zero,}
  JHEP {\bf 0911}, 086 (2009)
  [arXiv:0908.0056 [hep-th]].

\bibitem{DS} Dunford and Schwartz, {\it Linear Operators}, vol. I,II.

\bibitem{Polchinski} J.Polchinski, {\it String Theory. Volume I}, Cambridge University Press, Cambridge 1998.

\bibitem{Okawa02}
  Y.~Okawa,
  {\it Open string states and D-brane tension from vacuum string field theory,}
  JHEP {\bf 0207} (2002) 003
  [arXiv:hep-th/0204012].



 
\bibitem{Schwartz} L.~Schwartz, {\it Th\'eorie des distributions \`a valeurs vectorielles.I}
Ann.Inst.Fourier, 7 (1957) 1-141. {\it Th\'eorie des distributions \`a valeurs vectorielles.II}
Ann.Inst.Fourier, 8 (1958) 1-209. 


\bibitem{Gelfand} I.~M.~Guelfand and G.~E.~Chilov, {\it Les distributions}, tome 1. Dunod, Paris 1962.

\bibitem{MS} 
  M.~Murata and M.~Schnabl,
{\it Multibrane Solutions in Open String Field Theory,}
  arXiv:1112.0591 [hep-th].

\bibitem{EM2} 
  T.~Erler and C.~Maccaferri,
{\it Connecting Solutions in Open String Field Theory with Singular Gauge Transformations,}
  arXiv:1201.5119 [hep-th].


\bibitem{EM3} 
  T.~Erler and C.~Maccaferri,
{\it The Phantom Term in Open String Field Theory,}
  arXiv:1201.5122 [hep-th].


\end{thebibliography}
\end{document}